\documentclass[twocolumn]{jime}
\usepackage[latin1]{inputenc}
\usepackage[english]{babel}
\usepackage{graphicx}
\usepackage{amsmath}
\usepackage{amssymb}
\usepackage{color}
\usepackage{subfigure}

\newtheorem{remark}{Remark}

\begin{document}

\title{Volatility made observable at last}
\author{Michel \textsc{Fliess}\textsuperscript{1},
C\'{e}dric \textsc{Join}\textsuperscript{2,3}, Fr\'{e}d\'{e}ric \textsc{Hatt}\textsuperscript{4}\\
\normalsize \vskip 1em \textsuperscript{1}LIX (CNRS, UMR 7161),
\'Ecole polytechnique, 91228 Palaiseau, France\\
 \textit{Michel.Fliess@polytechnique.edu}
\vskip 1em \textsuperscript{2}CRAN (CNRS, UMR 7039),
Nancy-Universit\'e,
BP 239, 54506 Vand\oe uvre-l\`es-Nancy, France\\
\textit{Cedric.Join@cran.uhp-nancy.fr} \vskip 1em
\textsuperscript{3}\'Equipe Non-A, INRIA  Lille -- Nord-Europe,
France
\vskip 1em \textsuperscript{4} Lucid Capital Management, 2
avenue
Charles de Gaulle, BP 351, 2013 Luxembourg, Luxembourg \\
 \textit{hatt@lucid-cap.com}}
\maketitle

{\small\noindent{\small \textit{Abstract}--- The Cartier-Perrin
theorem, which was published in 1995 and is expressed in the
language of \emph{nonstandard analysis}, permits, for the first time
perhaps, a clear-cut mathematical definition of the
\emph{volatility} of a financial asset. It yields as a byproduct a
new understanding of the means of returns, of the beta coefficient,
and of the Sharpe and Treynor ratios. New estimation techniques from
automatic control and signal processing, which were already
successfully applied in quantitative finance, lead to several
computer experiments with some quite convincing forecasts.

 }}

\section{Introduction}
Although \emph{volatility}, which reflects the price fluctuations,
is ubiquitous in quantitative finance (see, \textit{e.g.},
\cite{bodie,franke,hull,roncalli,sinclair,wilmott}, and the
references therein), Paul
Wilmott writes rightly (\cite{wilmott}, chap. 49, p. 813): \\
\textit{Quite frankly, we do not know what volatility currently is,
never mind what it may be in the future}.\\ Our title is explained
by sentences like the following one in Tsay's book (\cite{tsay}, p. 98): \\
\textit{\dots volatility is not directly observable \dots}  \\ The
lack moreover of any precise mathematical definition leads to
multiple ways for computing volatility which are by no means
equivalent and might even be sometimes misleading (see,
\textit{e.g.}, \cite{gol}). Our theoretical formalism and the
corresponding computer simulations will confirm what most
practitioners already know. It is well
expressed by Gunn (\cite{gunn}, p. 49): \\
\textit{Volatility is not only referring to something that
fluctuates sharply up and down but is also referring to something
that moves sharply in a sustained direction}. \vskip 0.25cm The
existence of trends \cite{fes} for time series, which should be
viewed as the \emph{means}, or \emph{averages}, of those series,
yields
\begin{itemize}
\item a natural and straightforward model-free definition of the variance (resp.
covariance) of one (resp. two) time series,
\item simple forecasting techniques which are based on similar techniques to those
in \cite{fes,malo2,beta,delta}.
\end{itemize}
Exploiting the above approach to volatility for the return of some
financial asset necessitates some care due to the highly fluctuating
character of returns. This is accomplished by considering the means
of the time series associated to the prices logarithms. The
following results are derived as byproducts:
\begin{enumerate}
\item We complete \cite{beta} with a new definition of the classic
beta coefficient for returns. It should bypass most of the existing
criticisms.
\item The Sharpe (\cite{sharpe1,sharpe2}) and Treynor ratios, which are famous
performance measures for trading strategies (see, \textit{e.g.},
\cite{bodie,roncalli,treynor,wilmott}, and the references therein),
are connected to a quite arbitrary financial time series. They might
lead to new and useful trading {\em indicators}.
\end{enumerate}
\begin{remark}
The graphical representation of all the above quantities boils down
to the drawing of means which has been already successfully achieved
in \cite{fes,malo2,beta,delta}.
\end{remark}

\vskip 0.25cm

Our paper is organized as follows. After recalling the
Cartier-Perrin theorem \cite{cartier}, Section \ref{basics} defines
(co)variances and volatility. In order to apply this setting to
financial returns, Section \ref{ret} defines the means of returns
and suggests definitions of the beta coefficient, and of the Sharpe
and Treynor ratios. Numerous quite convincing computer experiments
are shown in Section \ref{computer}, which displays also excellent
forecasts for the volatility. Some short discussions on the concept
of volatility may be found in Section \ref{conclusion}.

\section{Mean, variance and covariance revisited}\label{basics}
\subsection{Time series via nonstandard analysis}
\subsubsection{Infinitesimal sampling} Take the time
interval $[0, 1] \subset \mathbb{R}$ and introduce as often in
\emph{nonstandard analysis} the infinitesimal sampling
$${\mathfrak{T}} = \{ 0 = t_0 < t_1 < \dots < t_\nu = 1 \}$$
where $t_{i+1} - t_{i}$, $0 \leq i < \nu$, is {\em infinitesimal},
{\it i.e.}, ``very small''.\footnote{See, \textit{e.g.},
\cite{diener1,diener2} for basics in nonstandard analysis.} A
\emph{time series} $X(t)$ is a function $X: {\mathfrak{T}}
\rightarrow \mathbb{R}$.

\subsubsection{$S$-integrability}
The {\em Lebesgue measure} on ${\mathfrak{T}}$ is the function $m$
defined on ${{\mathfrak{T}}} \backslash \{1\}$ by $m(t_{i}) =
t_{i+1} - t_{i}$. The measure of any interval $[c, d[ \subset
\mathfrak{I}$, $c \leq d$, is its length $d -c$.  The
\emph{integral} over $[c, d[$ of the time series $X(t)$ is the sum
$$\int_{[c, d[} Xdm = \sum_{t \in [c, d[} X(t)m(t)$$
$X$ is said to be $S$-{\em integrable} if, and only if, for any
interval $[c, d[$ the integral $\int_{[c, d[} |X| dm$ is
\emph{limited}\footnote{A real number is \emph{limited} if, and only
if, it is not infinitely large.} and, if $d - c$ is infinitesimal,
also infinitesimal.
\subsubsection{Continuity and Lebesgue integrability}
$X$ is $S$-{\em continuous} at $t_\iota \in {\mathfrak{T}}$ if, and
only if, $f(t_\iota) \simeq f(\tau)$ when $t_\iota \simeq
\tau$.\footnote{$a \simeq b$ means that $a - b$ is infinitesimal.}
$X$ is said to be {\em almost continuous} if, and only if, it is
$S$-continuous on ${\mathfrak{T}} \setminus R$, where $R$ is a {\em
rare} subset.\footnote{The set $R$ is said to be \emph{rare}
\cite{cartier} if, for any standard real number $\alpha > 0$, there
exists an internal set $B \supset A$ such that $m(B) \leq \alpha$.}
$X$ is \emph{Lebesgue integrable} if, and only if, it is
$S$-integrable and almost continuous.
\subsubsection{Quick fluctuations}
A time series ${\mathcal{X}}: {\mathfrak{T}} \rightarrow \mathbb{R}$
is said to be {\em quickly fluctuating}, or {\em oscillating}, if,
and only if, it is $S$-integrable and $\int_A {\mathcal{X}} dm$ is
infinitesimal for any {\em quadrable} subset.\footnote{A set is
\emph{quadrable} \cite{cartier} if its boundary is rare.}
\subsubsection{The Cartier-Perrin theorem}

Let $X: {\mathfrak{T}} \rightarrow \mathbb{R}$ be a $S$-integrable
time series. Then, according to the Cartier-Perrin theorem
\cite{cartier},\footnote{Remember that this result led to a new
foundation \cite{bruit} of the analysis of noises in automatic
control and in signal processing. A more down to earth exposition
may be found in \cite{lobry}.} the additive decomposition
\begin{equation}\label{decomposition}
\boxed{X(t) = E(X)(t) + X_{\tiny{\rm fluctuation}}(t)}
\end{equation}
holds where
\begin{itemize}
\item the \emph{mean}, or \emph{average}, $E(X)(t)$ is Lebesgue
integrable,\footnote{$E(X)(t)$ was called \emph{trend} in our
previous publications \cite{fes,malo2,beta,delta}.}
\item $X_{\tiny{\rm fluctuation}}(t)$ is quickly fluctuating.
\end{itemize}
The decomposition \eqref{decomposition} is unique up to an
infinitesimal.

\begin{remark}$E(X)(t)$, which is ``smoother'' than $X(t)$, provides a
mathematical justification \cite{fes} of the \emph{trends} in
\emph{technical analysis} (see, \textit{e.g.}, \cite{bechu,kirk}).
\end{remark}

\begin{remark}
Calculations of the means and of its derivatives, if they exist, are
deduced, via new estimation techniques, from the denoising results
in \cite{nl,mboup} (see also \cite{gretsi}), which extend the
familiar moving averages, which are classic in technical analysis
(see, \textit{e.g.}, \cite{bechu,kirk}).
\end{remark}

\subsection{Variances and covariances}
\subsubsection{Squares and products}
Take two $S$-integrable time series $X(t)$, $Y(t)$, such that their
squares and the squares of $E(X)(t)$ and $E(Y)(t)$ are also
$S$-integrable. The Cauchy-Schwarz inequality shows that the
products
\begin{itemize}
\item $X(t)Y(t)$, $E(X)(t)E(Y)(t)$,
\item $E(X)(t)Y_{\tiny{\rm fluctuation}}(t)$,
$X_{\tiny{\rm fluctuation}}(t)E(Y)(t)$,
\item $X_{\tiny{\rm
fluctuation}}(t)Y_{\tiny{\rm fluctuation}}(t)$
\end{itemize}
are all $S$-integrable.
\subsubsection{Differentiability}\label{differentiability}
Assume moreover that $E(X)(t)$ and $E(Y)(t)$ are
\emph{differentiable} in the following sense: there exist two
Lebesgue integrable time series $f, g: {\mathfrak{T}} \rightarrow
\mathbb{R}$, such that, $\forall ~ t \in {\mathfrak{T}}$, with the
possible exception of a limited number of values of $t$, $E(X)(t) =
E(X)(0) + \int_{0}^{t} f(\tau)d\tau$, $E(Y)(t) = E(Y)(0) +
\int_{0}^{t} g(\tau)d\tau$. Integrating by parts shows that the
products $E(X)(t)Y_{\tiny{\rm fluctuation}}(t)$ and $X_{\tiny{\rm
fluctuation}}(t)E(Y)(t)$ are quickly fluctuating \cite{bruit}.

\begin{remark}
Let us emphasize that the product $$X_{\tiny{\rm
fluctuation}}(t)Y_{\tiny{\rm fluctuation}}(t)$$ is not necessarily
quickly fluctuating.
\end{remark}
\subsubsection{Definitions}
\begin{enumerate}
\item The \emph{covariance} of two time series $X(t)$ and $Y(t)$ is
\begin{eqnarray*}
\mbox{\rm cov}(XY)(t) & = & E\left((X - E(X))(Y - E(Y))
\right)(t) \\ & \simeq & E(XY)(t) - E(X)(t) \times E(Y)(t)
\end{eqnarray*}
\item The \emph{variance} of the time series $X(t)$ is
\begin{eqnarray*}\label{var}
\mbox{\rm var}(X)(t) & = & E\left((X - E(X))^2 \right)(t) \\ &
\simeq & E(X^2)(t) - \left(E(X)(t)\right)^2
\end{eqnarray*}
\item The \emph{volatility} of $X(t)$ is the corresponding standard
deviation
\begin{equation}\label{vol}
\boxed{\mbox{\rm vol}(X)(t) = \sqrt{\mbox{\rm var}(X)(t)}}
\end{equation}
\end{enumerate}
The volatility of a quite arbitrary time series seems to be
precisely defined here for the first time.
\begin{remark}
Another possible definition of the volatility (see \cite{gol}),
which is not equivalent to Equation \eqref{vol}, is the following
one
$$
E\left( | X - E(X)| \right)(t)
$$
It will not be exploited here.
\end{remark}
\section{Returns}\label{ret}
\subsection{Definition}
Assume from now on that, for any $t \in {\mathfrak{T}}$,
$$ 0 < m < X(t) < M$$
where $m$, $M$ are \emph{appreciable}.\footnote{A real number is
\emph{appreciable} if, and only if, it is neither infinitely small
nor infinitely large.} This is a realistic assumption if $X(t)$ is
the price of some financial asset $\mathfrak{A}$. The
\emph{logarithmic} \emph{return}, or \emph{log-return},\footnote{The
terminology \emph{continuously compounded return} is also used. See,
\textit{e.g.}, \cite{campbell} for more details.} of $X$ with
respect to some limited time interval $\Delta T > 0$ is the time
series $R_{\Delta T}$ defined by
\begin{equation*}\label{RR}
R_{\Delta T}(X) (t) = \ln \left( \frac{X(t)}{X(t - \Delta T)}
\right) = \ln X(t) - \ln X(t - \Delta T)
\end{equation*}
From $\frac{X(t)}{X(t - \Delta T)} = 1 + \frac{X(t) - X(t - \Delta
T)}{X(t - \Delta T)}$, we know that
\begin{equation}\label{RRR}
R_{\Delta T}(X) (t) \simeq \frac{X(t) - X(t - \Delta T)}{X(t -
\Delta T)}
\end{equation}
if $X(t) - X(t - \Delta T)$ is infinitesimal. The right handside of
Equation \eqref{RRR} is the \emph{arithmetic} return.

The \emph{normalized} logarithmic return is
\begin{equation}\label{nor}
r_{\Delta T}(X) (t) = \frac{R_{\Delta T} (t)}{\Delta T}
\end{equation}

\subsection{Mean}
\subsubsection{Definition} Replace $X: {\mathfrak{T}} \rightarrow
\mathbb{R}$ by
$$\ln X:
{\mathfrak{T}} \rightarrow {\mathbb{R}}, \quad t \mapsto \ln \left(
X(t) \right)$$ where the logarithms of the prices are taken into
account. Apply the Cartier-Perrin theorem to $\ln X$. The
\emph{mean}, or \emph{average}, of $r_{\Delta T}(t)$ given by
Equation \eqref{nor} is
\begin{equation}\label{aver}
\boxed{ \bar{r}_{\Delta T}(X) (t) = \frac{E (\ln X) (t) - E (\ln X)
(t - \Delta T)}{\Delta T}}
\end{equation}
As a matter of fact $r_{\Delta T}(X)$ and $\bar{r}_{\Delta T}(X)$
are related by
\begin{equation*}\label{avera}
r_{\Delta T}(X) (t) =  \bar{r}_{\Delta T}(X) (t) + \text{quick
fluctuations}
\end{equation*}
Assume that $E (X)$ and $E (\ln X)$ are differentiable according to
Section \ref{differentiability}. Call the derivative of $E (\ln X)$
the {\em normalized mean logarithmic instantaneous return} and write
\begin{equation}\label{r}
\boxed{ \bar{r}(X) (t) = \frac{d}{dt} E (\ln X) (t) }
\end{equation}
Note that $E (\ln X) (t) \simeq \ln \left( E (X) (t) \right) $ if in
Equation \eqref{decomposition} $X_{\tiny{\rm fluctuation}}(t) \simeq
0$. Then $\bar{r}(X) (t) \simeq \frac{\frac{d}{dt} E (X) (t)}{E (X)
(t)}$.
\subsubsection{Application to beta}\label{bet} Take two
assets $\mathfrak{A}$ and $\mathfrak{B}$ such that their normalized
logarithmic returns $r_{\Delta T} ({\mathfrak{A}})(t)$ and
$r_{\Delta T}({\mathfrak{B}}) (t)$, defined by Equation \eqref{nor},
exist.\footnote{This Section is adapting for returns the
presentation in \cite{beta}.} Following Equation \eqref{aver},
consider the space curve $t, \bar{r}_{\Delta T}({\mathfrak{A}}) (t),
\bar{r}_{\Delta T}({\mathfrak{B}}) (t)$ in the Euclidean space with
coordinates $t, x, y$. Its projection on the $x, y$ plane is the
plane curve $\mathfrak{C}$ defined by
$$
x_{\mathfrak{C}} (t) = \bar{r}_{\Delta T}({\mathfrak{A}}) (t),
y_{\mathfrak{C}} (t) = \bar{r}_{\Delta T}({\mathfrak{B}}) (t)
$$
The tangent of $\mathfrak{C}$ at a regular point, which is defined
by $\frac{d x_{\mathfrak{C}} (t)}{dt}, \frac{d y_{\mathfrak{C}}
(t)}{dt}$, yields, if $\frac{d x_{\mathfrak{C}} (t)}{dt} \neq 0$,
\begin{equation}\label{inc}
\Delta y_{\mathfrak{C}}  \approx \beta (t) \Delta x_{\mathfrak{C}}
\end{equation}
where
\begin{itemize}
\item $\Delta x_{\mathfrak{C}} = x_{\mathfrak{C}} (t + h) -
x_{\mathfrak{C}} (t)$, $\Delta y_{\mathfrak{C}} = y_{\mathfrak{C}}
(t + h) - y_{\mathfrak{C}} (t)$;
\item $h \in \mathbb{R}$ is ``small'';
\item
\begin{equation}\label{beta1} {\beta
(t) = \frac{\frac{d y_{\mathfrak{C}} (t)}{dt}}{\frac{d
x_{\mathfrak{C}} (t)}{dt}}}
\end{equation}
\end{itemize}
When $y_{\mathfrak{C}} (t)$ may be viewed locally as a smooth
function of $x_{\mathfrak{C}} (t)$, Equation \eqref{beta1} becomes
\begin{equation*}\label{beta2} {\beta
(t) = \frac{d y_{\mathfrak{C}}}{dx_{\mathfrak{C}}}}
\end{equation*}
\subsubsection{The Treynor ratio of an asset}
Let $\beta_{\mathfrak{M}} ({\mathfrak{A}}) (t)$ be the beta
coefficient defined in Section \ref{bet} for ${\mathfrak{A}}$ with
respect to the market portfolio $\mathfrak{M}$. Define the
\emph{Treynor ratio} and the \emph{instantaneous Treynor ratio} of
$\mathfrak{A}$ with respect to $\mathfrak{M}$ respectively by
\begin{equation*}\label{treynor}
\boxed{\text{TR}_{{\mathfrak{M}}, \Delta T}({\mathfrak{A}}) (t) =
\frac{\bar{r}_{\Delta T}({\mathfrak{A}}) (t)}{\beta_{\mathfrak{M}}
({\mathfrak{A}}) (t)},  \quad
\text{TR}_{{\mathfrak{M}}}({\mathfrak{A}}) (t) =
\frac{\bar{r}({\mathfrak{A}}) (t)}{\beta_{\mathfrak{M}}
({\mathfrak{A}}) (t)}}
\end{equation*}

\subsection{Volatility}
Formulae \eqref{vol}, \eqref{nor}, \eqref{aver}, \eqref{r} yield the
following mathematical definition of the \emph{volatility} of the
asset $\mathfrak{A}$:
\begin{equation}\label{vola}
\boxed{{\text{\bf vol}}_{\Delta T}   ({\mathfrak{A}})(t)  = \sqrt{E
(r_{\Delta T} - \bar{r}_{\Delta T})^2  (t)}}
\end{equation}
which yields
$$
\boxed{{\text{\bf vol}}_{\Delta T}  ({\mathfrak{A}})(t) \simeq
\sqrt{E(r_{\Delta T}^2)(t) - (\bar{r}_{\Delta T} (t))^2}}
$$
The value at time $t$ of ${\text{\bf vol}}_{\Delta T}
({\mathfrak{A}})$ may be viewed as the \emph{actual} volatility
(see, \textit{e.g.}, \cite{wilmott}, chap. 49, pp. 813-814).
\begin{remark}\label{fondamental}
A crucial difference between Formula \eqref{vola} and the usual
\emph{historical}, or \emph{realized}, volatilities (see,
\textit{e.g.}, \cite{wilmott}, chap. 49, pp. 813-814) lies in the
presence of a non-constant mean. It is often assumed to be $0$ in
the existing literature.
\end{remark}
\begin{remark}There is no connection with
\begin{itemize}
\item the \emph{implied} volatility, which
is connected to the Black-Scholes modeling (see, \textit{e.g.},
\cite{wilmott}, chap. 49, pp. 813-814),
\item the recent {\em model-free implied volatility} (see \cite{mfv},
and \cite{jiang,rouah}), although the origin of our viewpoint may be
partly traced back to our \emph{model-free control} strategy
(\cite{esta,edf}).
\end{itemize}
\end{remark}

\subsection{The Sharpe ratio of an asset}
Define the \emph{Sharpe ratio} of the asset ${\mathfrak{A}}$ by
\begin{equation}\label{sharpe}
\boxed{\text{SR}_{\Delta T}({\mathfrak{A}}) (t) =
\frac{\bar{r}_{\Delta T}({\mathfrak{A}}) (t)}{{\text{\bf
vol}}_{\Delta T}({\mathfrak{A}}) (t) }}
\end{equation}
According to \cite{ald}, p. 52, it is quite close to some
utilization of the Sharpe ratio in high-frequency trading.

\section{Computer experiments}\label{computer}

\begin{figure*}
\center\subfigure[Daily price]{\rotatebox{-0}{\includegraphics*[width=0.88\columnwidth]{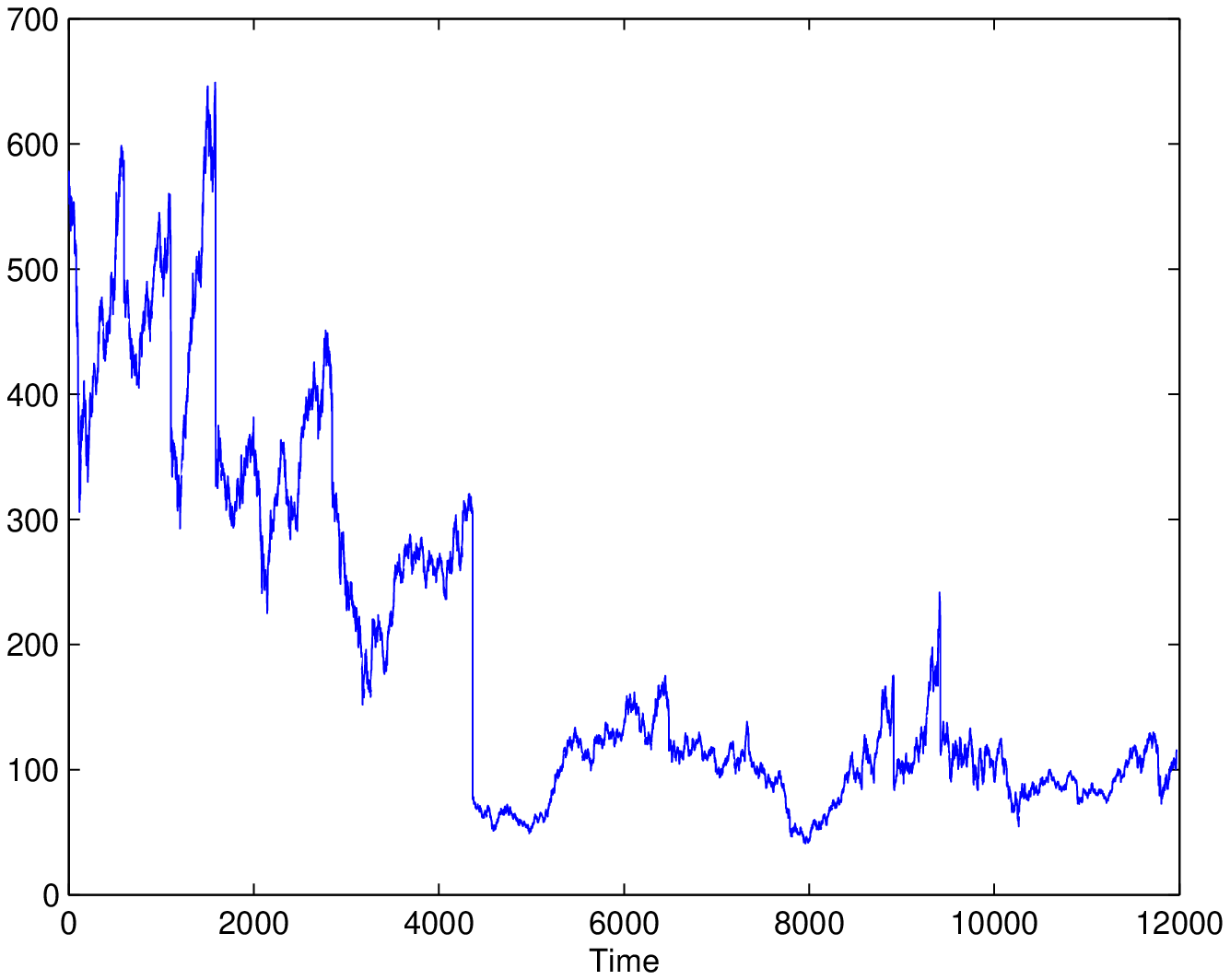}}}\\
\subfigure[Normalized logarithmic return
$r(t)$]{\rotatebox{-0}{\includegraphics*[width=
0.88\columnwidth]{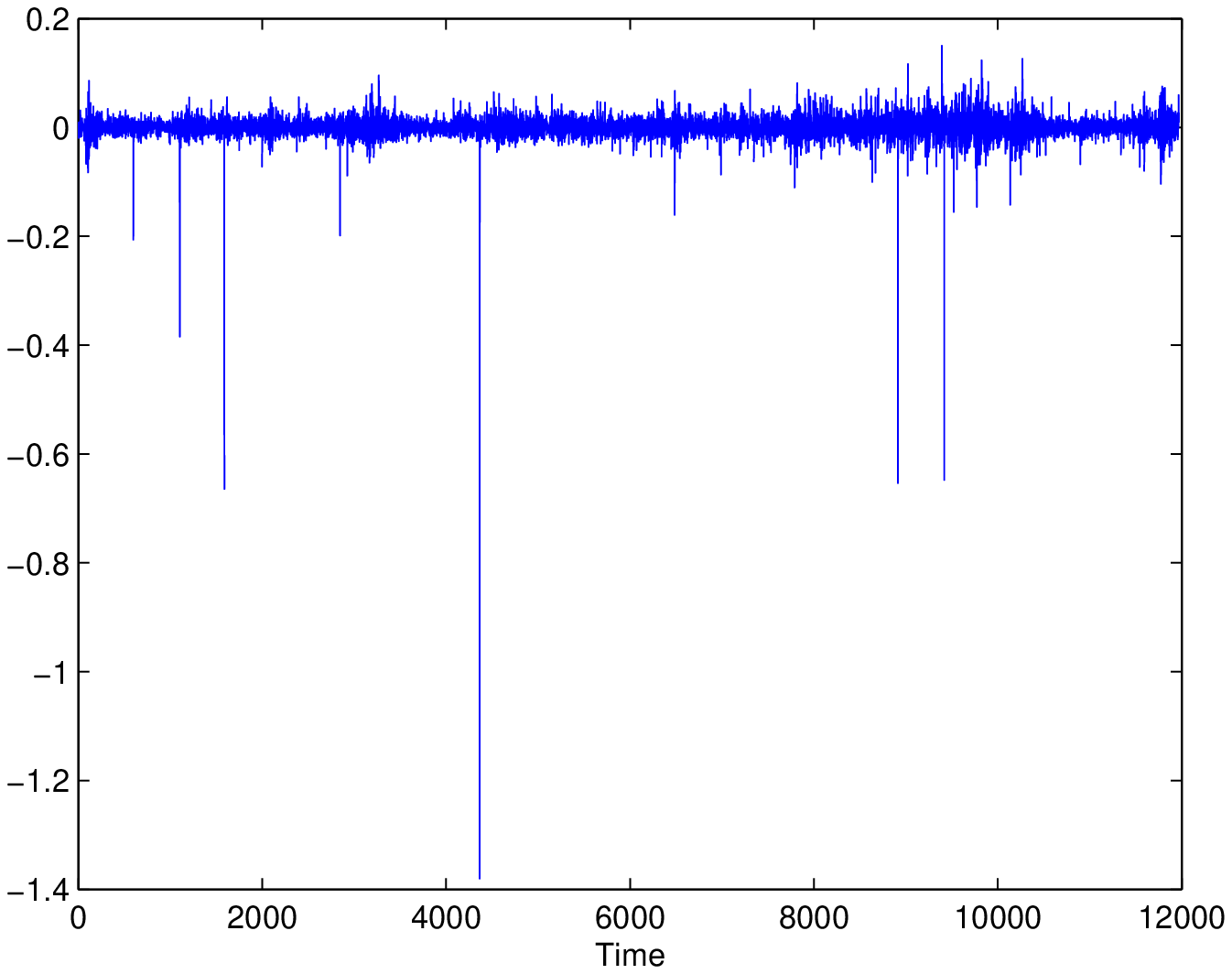}}}
\subfigure[Normalized mean logarithmic return $\bar
r(t)$]{\rotatebox{-0}{\includegraphics*[width=
0.88\columnwidth]{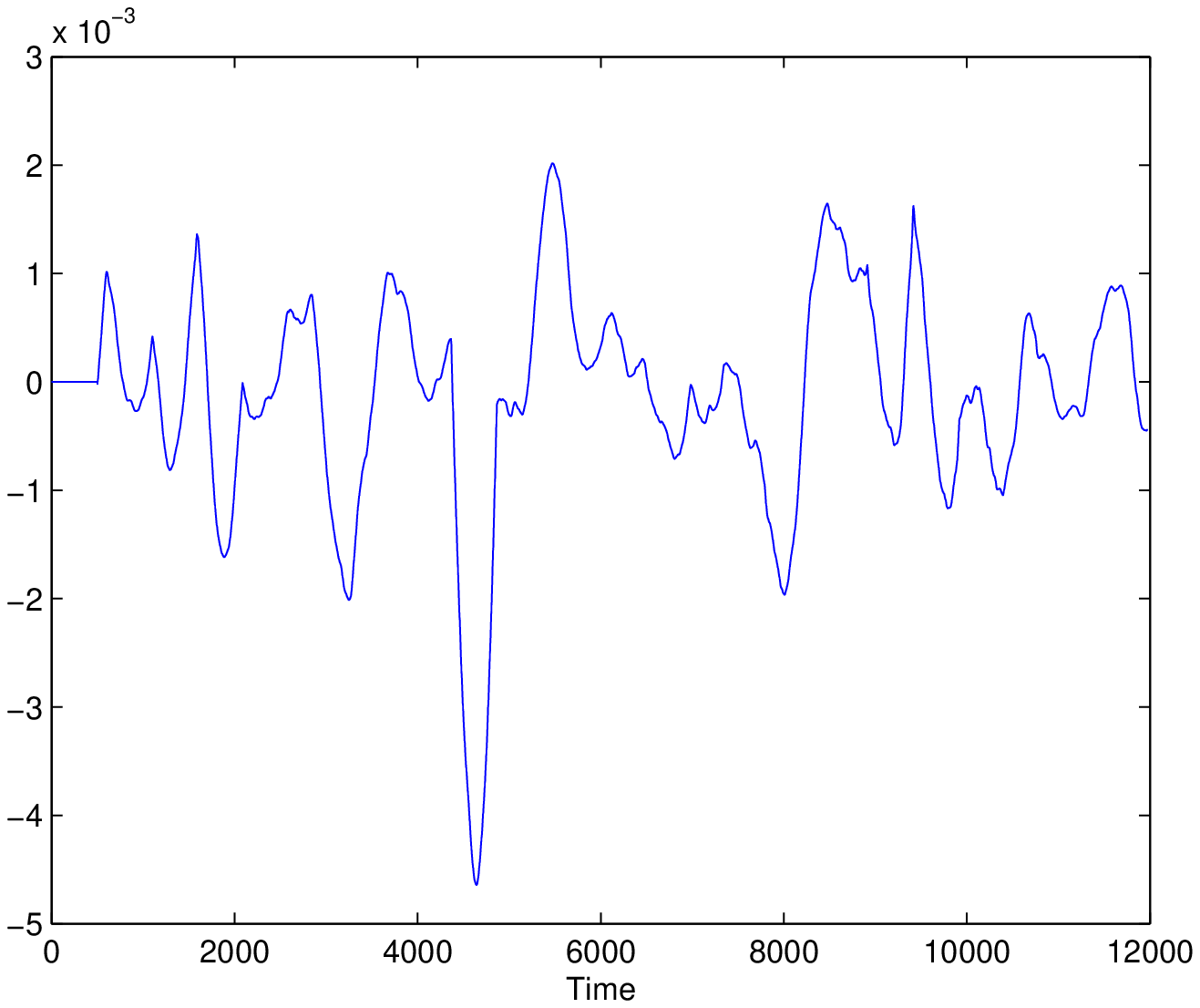}}}
\subfigure[$\text{\bf vol}({IBM})(t)$ (--) and 5 days forecasting (-
-)]{\rotatebox{-0}{\includegraphics*[width=
0.88\columnwidth]{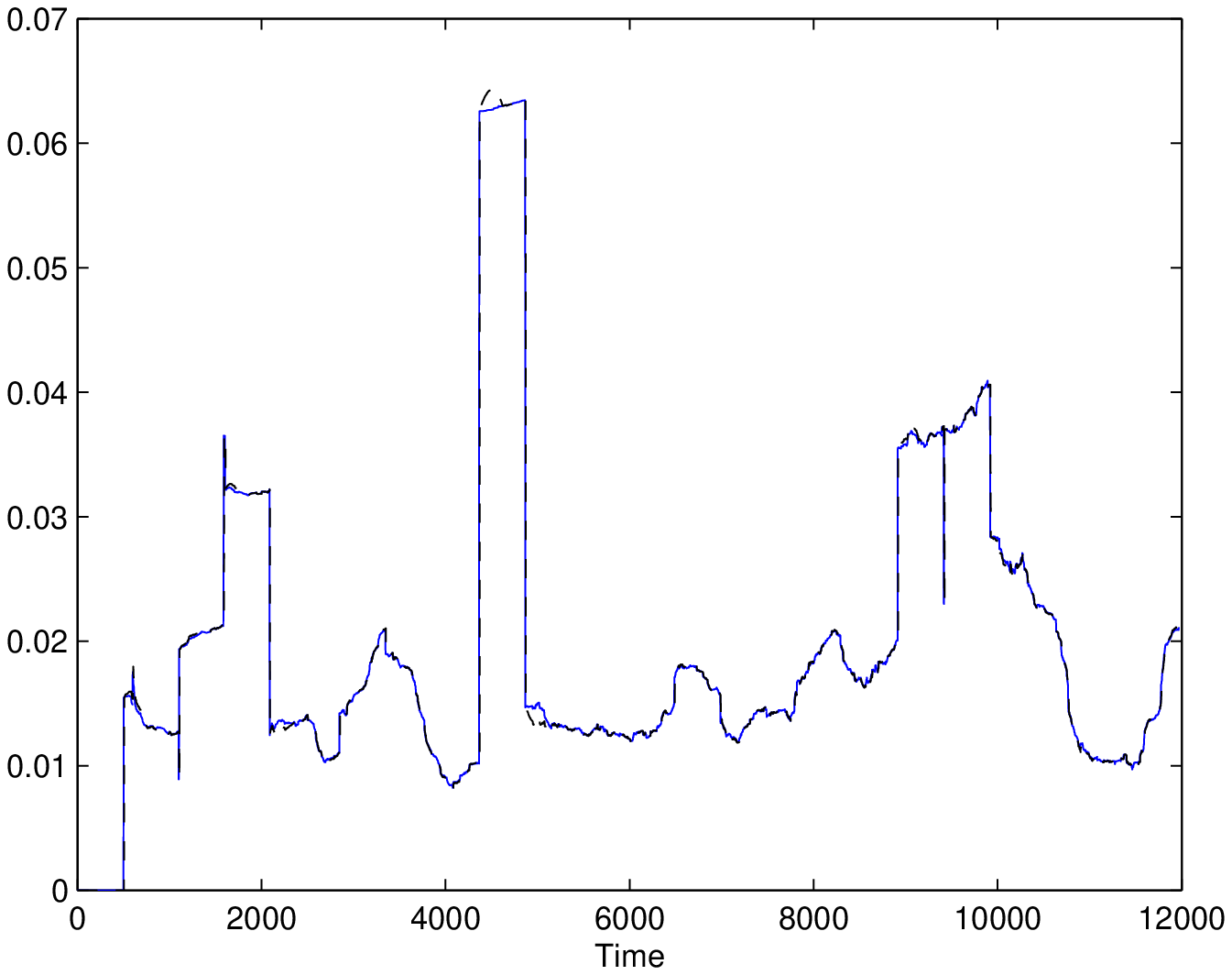}}}
\subfigure[$\text{\bf vol}({IBM})(t)$ (--) and 20 days forecasting
(- -)]{\rotatebox{-0}{\includegraphics*[width=
0.88\columnwidth]{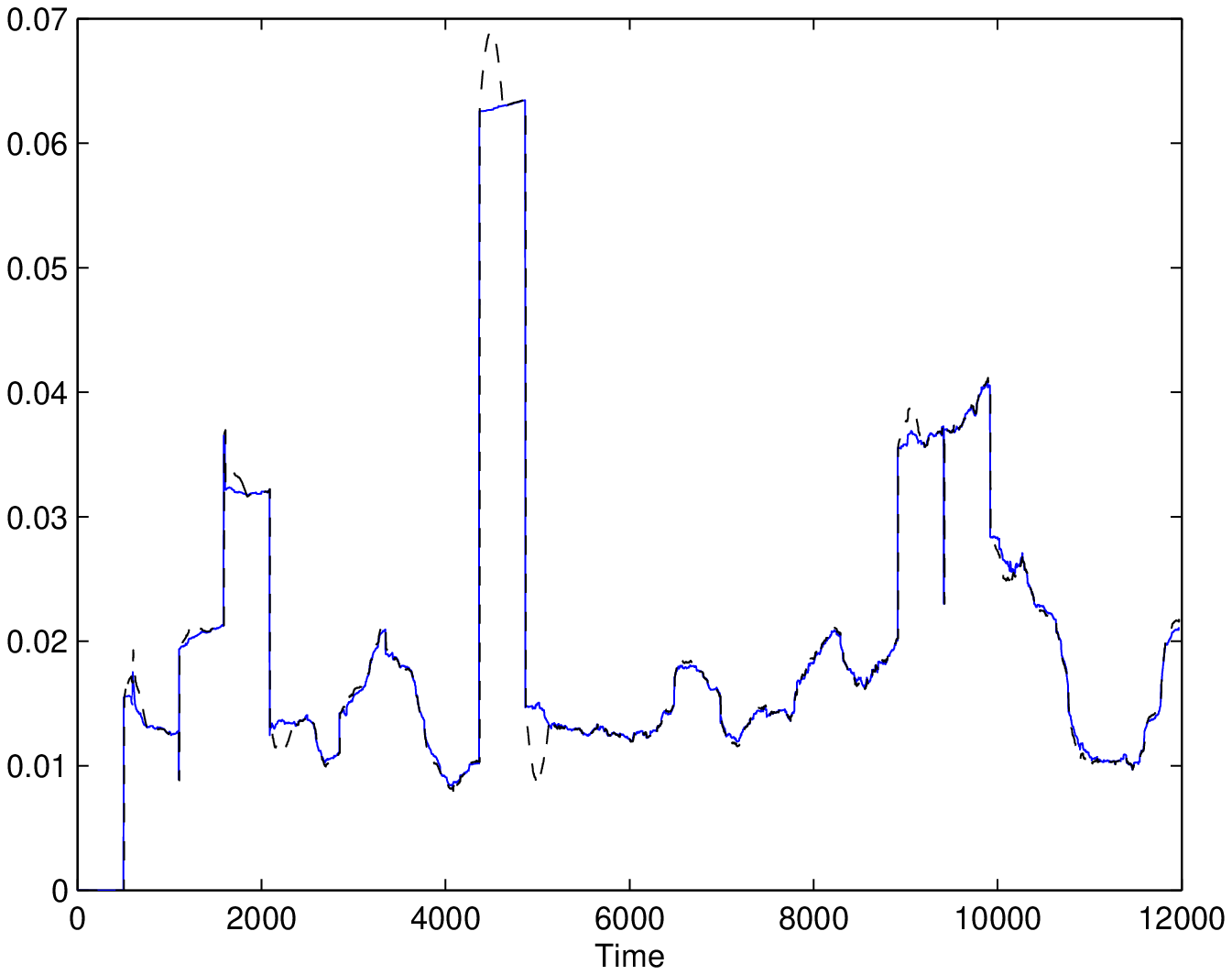}}}
 \caption{IBM \label{IBM}}
\end{figure*}

\begin{figure*}
\center\subfigure[Modified normalized logarithmic return
$r(t)$]{\rotatebox{-0}{\includegraphics*[width=
0.88\columnwidth]{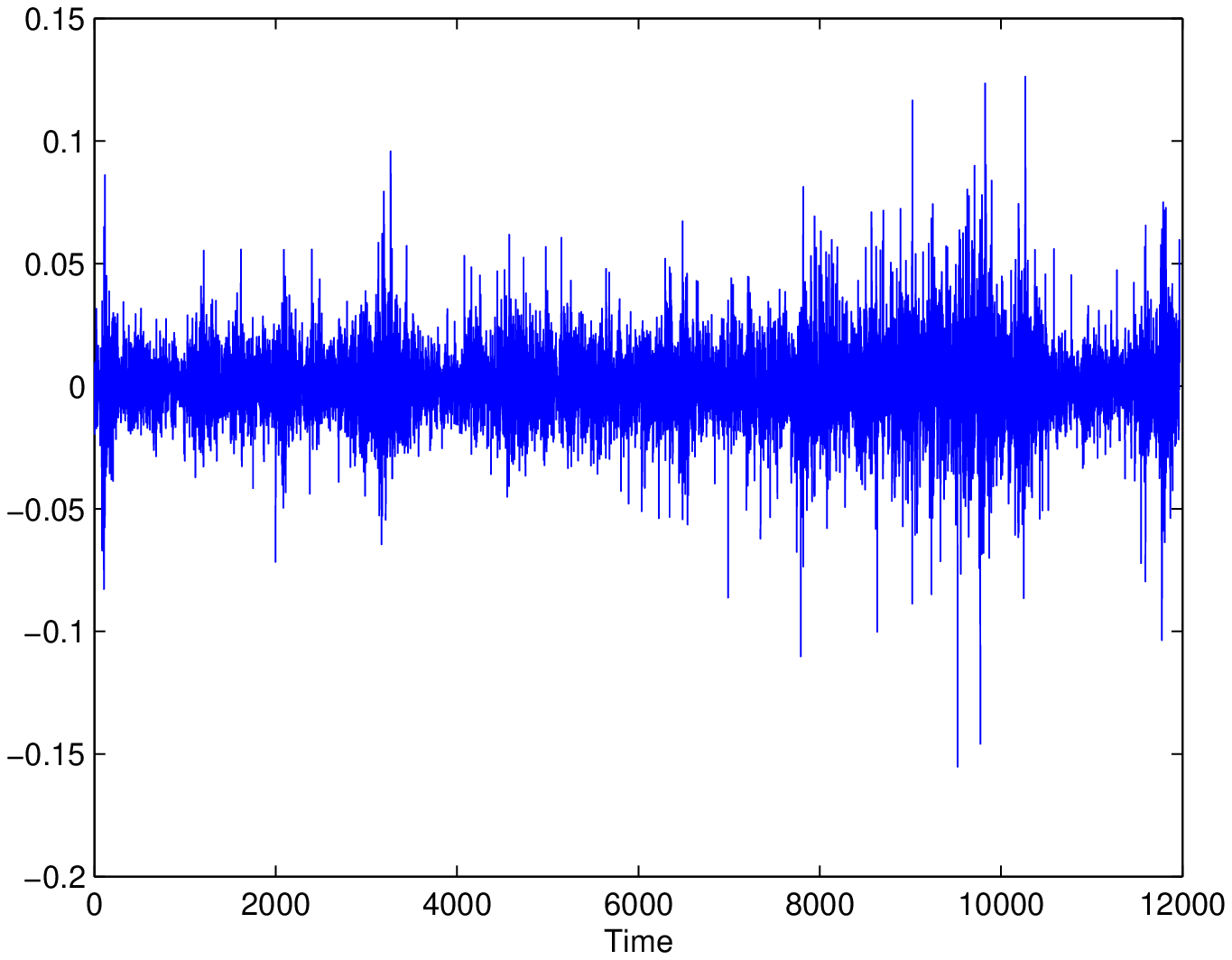}}}
\subfigure[$\text{\bf vol}({IBM})(t)$ (--) and 20 days forecasting
(- -)]{\rotatebox{-0}{\includegraphics*[width=
0.88\columnwidth]{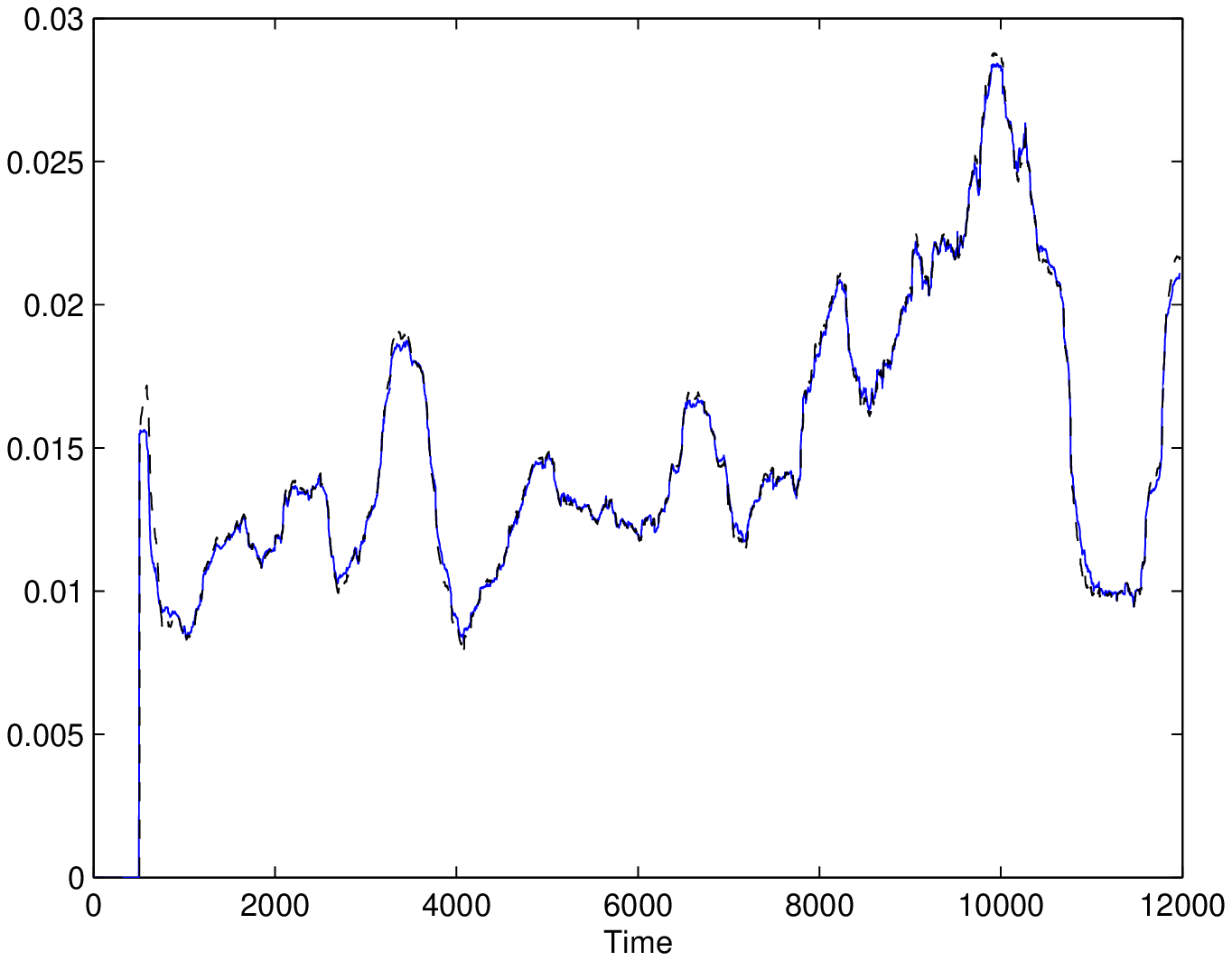}}}
 \caption{IBM\label{IBMsuite}}
\end{figure*}

\begin{figure*}
\center\subfigure[Daily
price]{\rotatebox{-0}{\includegraphics*[width=0.88\columnwidth]{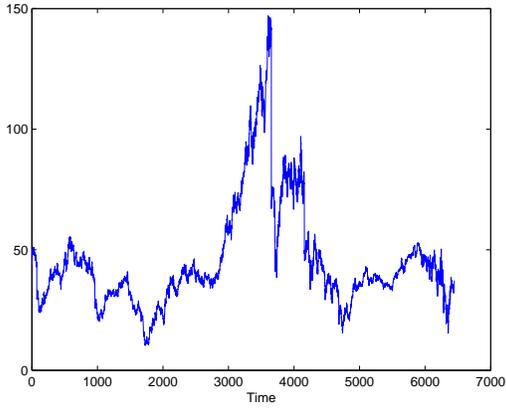}}}
\subfigure[Modified normalized logarithmic return
$r(t)$]{\rotatebox{-0}{\includegraphics*[width=0.88\columnwidth]{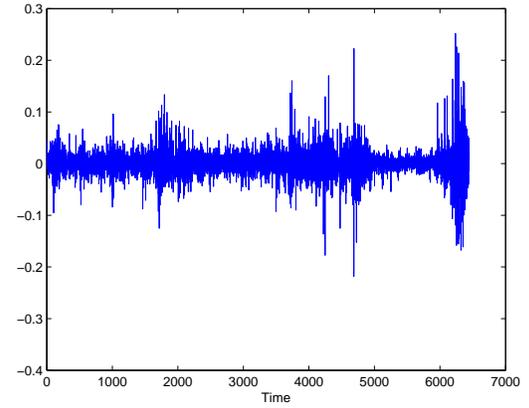}}}
\subfigure[Normalized mean logarithmic return $\bar
r(t)$]{\rotatebox{-0}{\includegraphics*[width=
0.88\columnwidth]{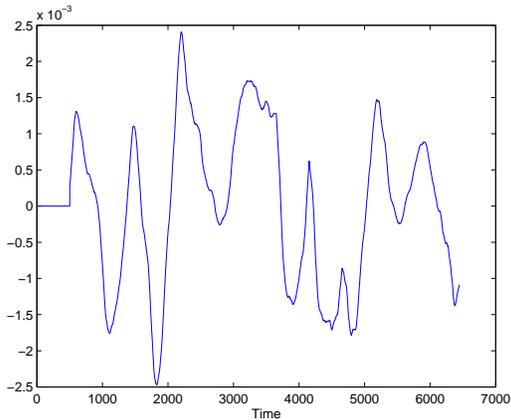}}}
\subfigure[$\text{\bf vol}({JPM})(t)$ (--) and 20 days forecasting
(- -)]{\rotatebox{-0}{\includegraphics*[width=
0.88\columnwidth]{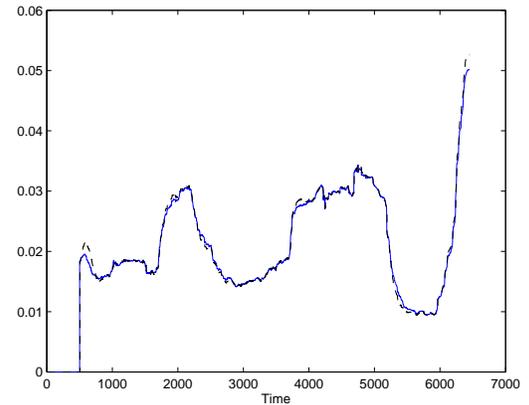}}}
 \caption{JPMORGAN CHASE (JPM) \label{JPM}}
\end{figure*}

\begin{figure*}
\center\subfigure[Daily
price]{\rotatebox{-0}{\includegraphics*[width=0.88\columnwidth]{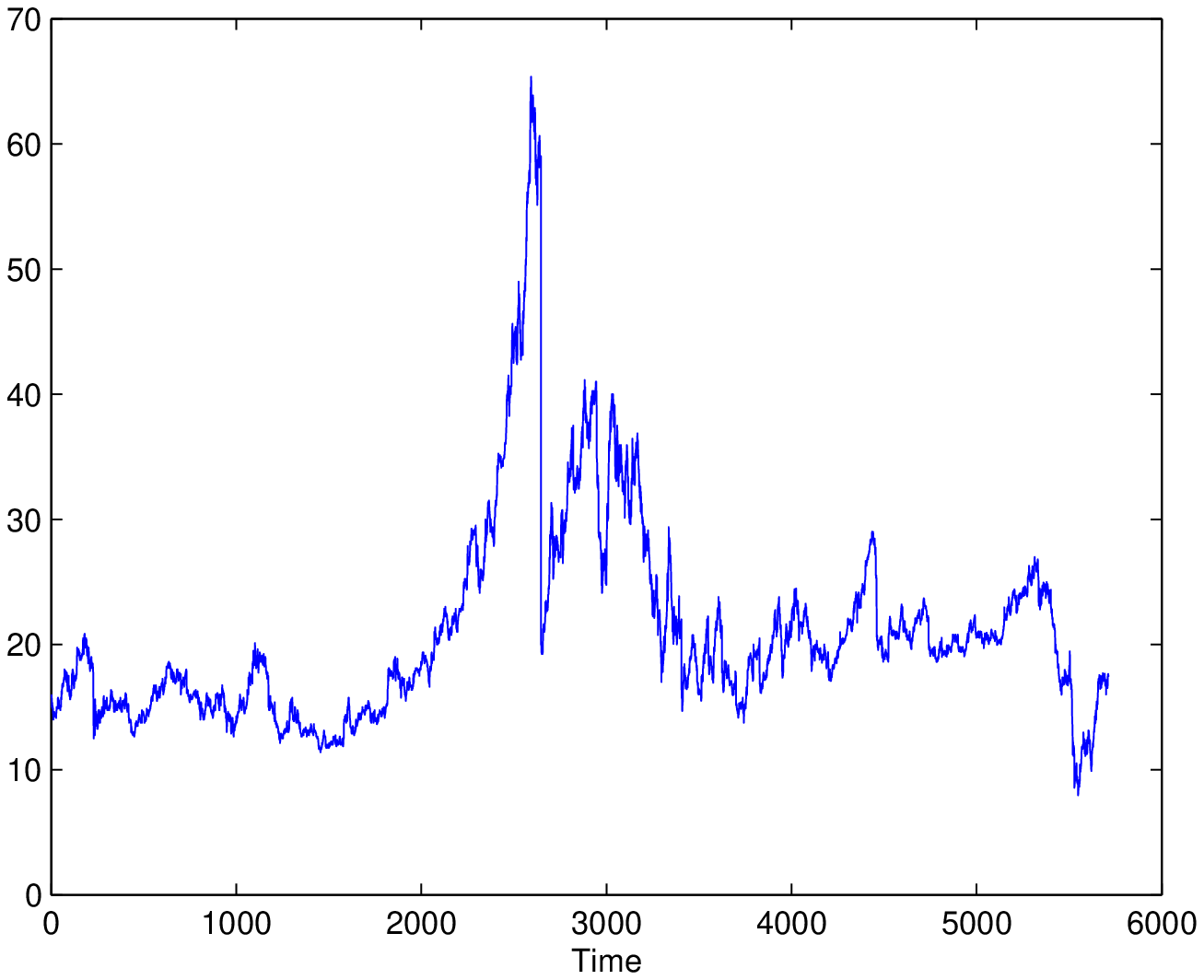}}}
\subfigure[Modified normalized logarithmic return
$r(t)$]{\rotatebox{-0}{\includegraphics*[width=0.88\columnwidth]{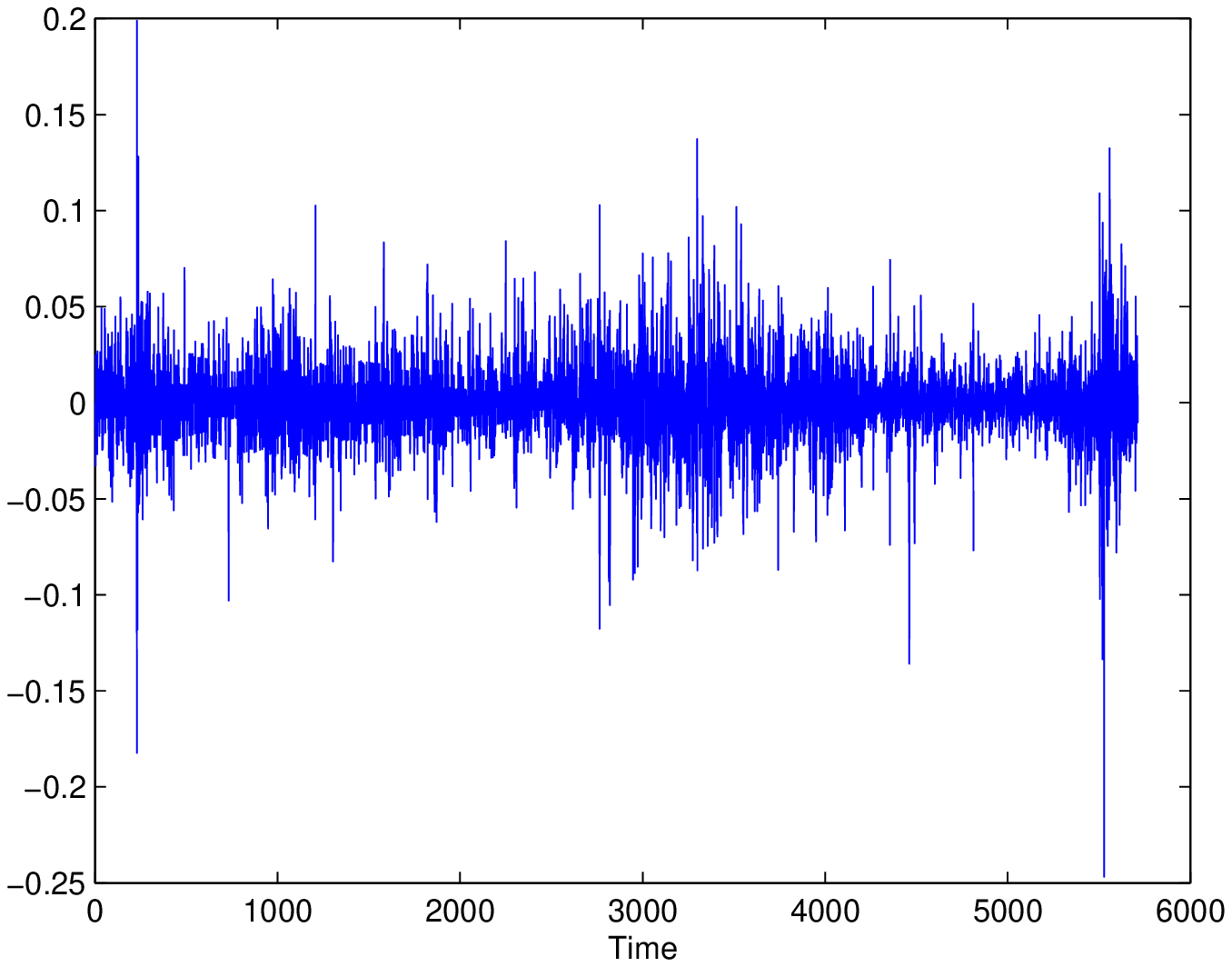}}}
\subfigure[Normalized mean logarithmic return $\bar
r(t)$]{\rotatebox{-0}{\includegraphics*[width=
0.88\columnwidth]{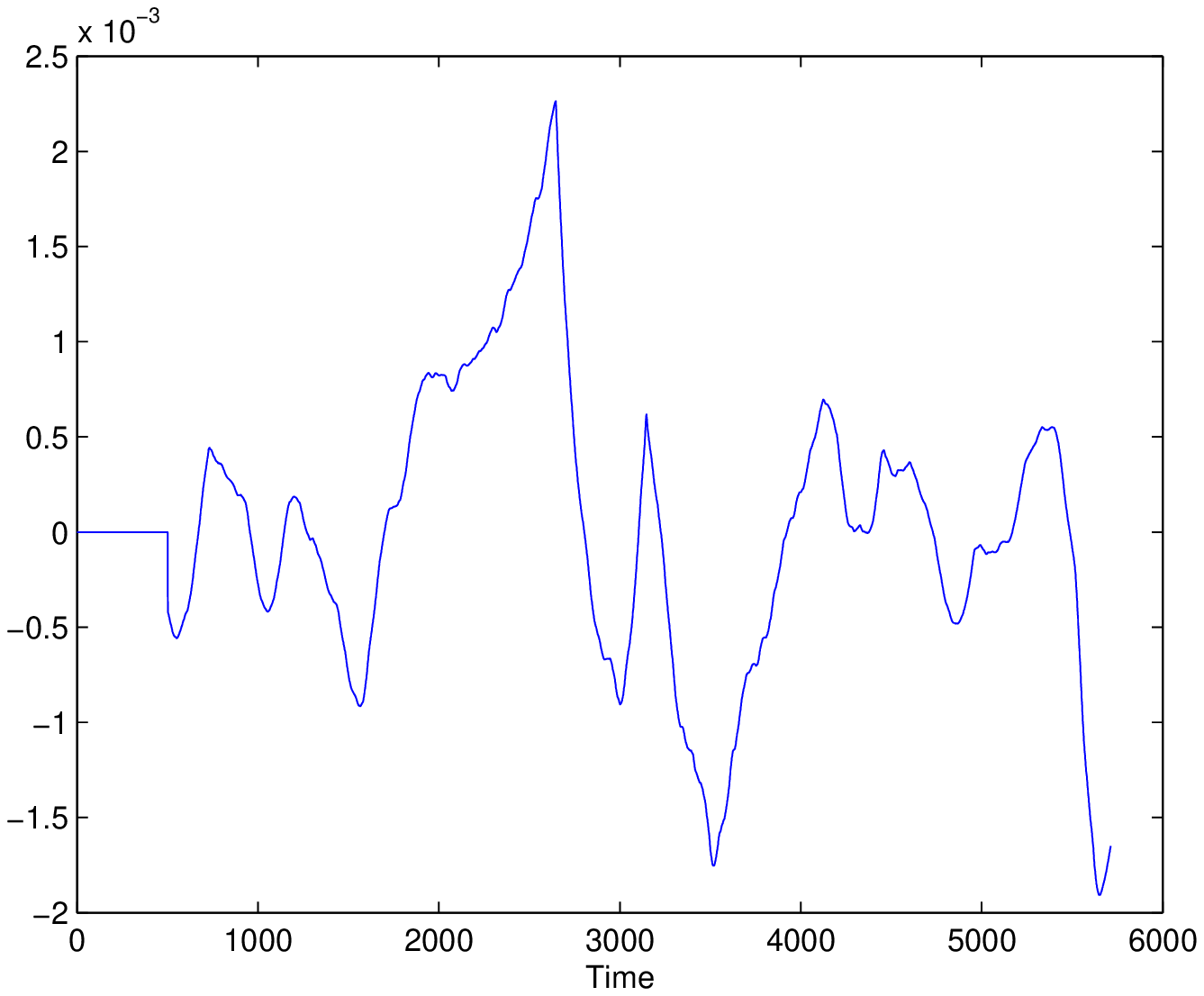}}}
\subfigure[$\text{\bf vol}({CCE})(t)$ (--) and 20 days forecasting
(- -)]{\rotatebox{-0}{\includegraphics*[width=
0.88\columnwidth]{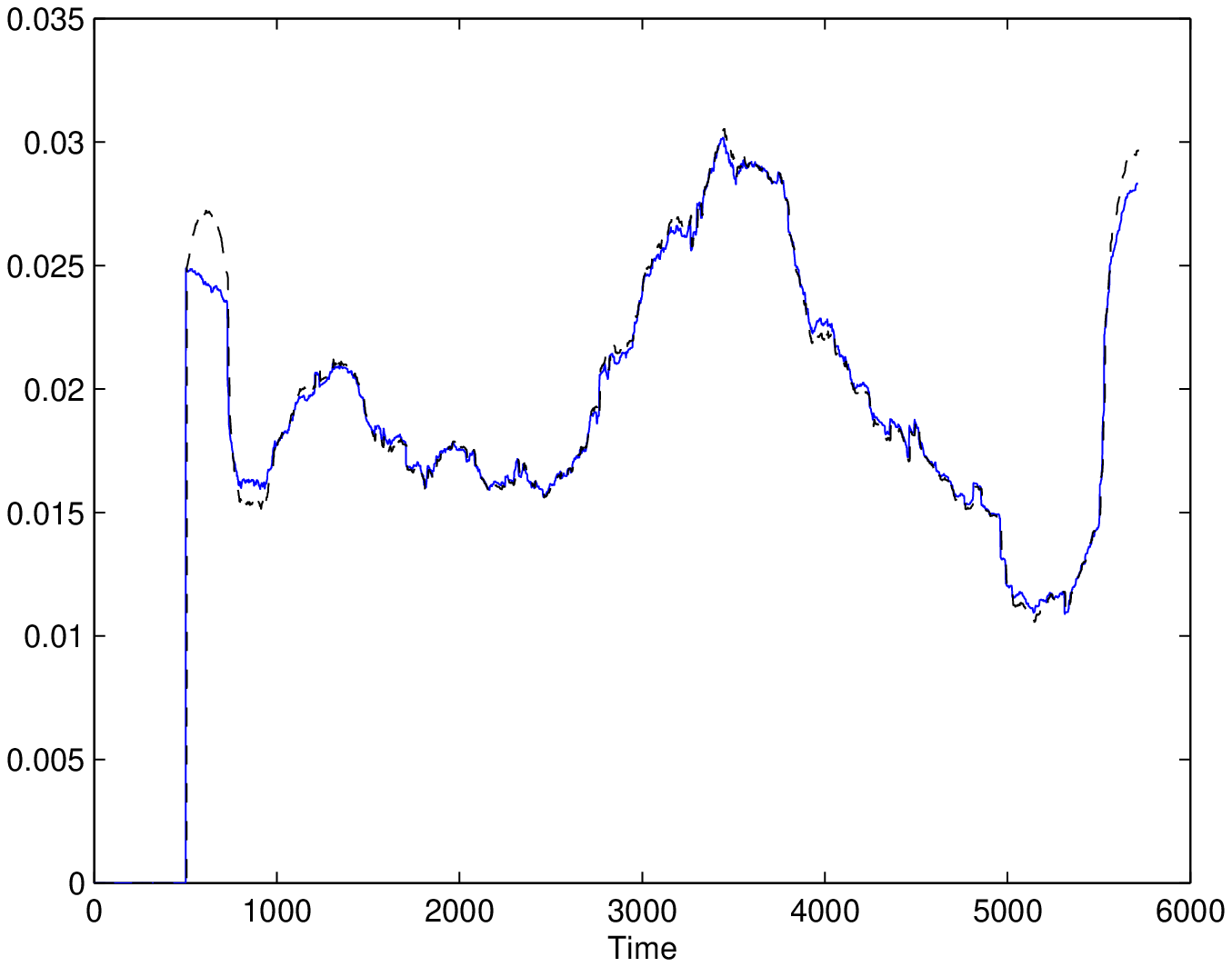}}}
 \caption{COCA COLA (CCE) \label{CCE}}
\end{figure*}

We have utilized the following three listed shares:
\begin{enumerate}
\item IBM from 1962-01-02 to 2009-07-21 (11776 days) (Figures \ref{IBM} and
\ref{IBMsuite}),
\item JPMORGAN CHASE (JPM) from 1983-12-30 until 2009-07-21 (6267 days) (Figures
\ref{JPM}),
\item COCA COLA (CCE) from 1986-11-24 until 2009-07-21 (5519 days) (Figures
\ref{CCE}).
\end{enumerate}
Figures \ref{IBM} and \ref{JPM} show a ``better'' behavior for the
normalized mean logarithmic return \eqref{r}, \textit{i.e.},
$\bar{r}(t)$ is less affected by an abrupt short price variation.
Such variations are nevertheless causing important variations on our
volatility, with only a ``slow mean return''. We suggest an adaptive
threshold for attenuating this annoying feature, which does not
reflect well the price behavior. Note the excellent volatility
forecasts which are obtained via elementary numerical recipes as in
\cite{fes,malo2,beta,delta}. Our forecasting results, which are
easily computable, seem to be more reliable than those obtained via
the celebrated ARCH type techniques, which go back to Engle (see
\cite{engle} and the references therein).\footnote{Those comparisons
need to be further investigated.}

\begin{figure*}
\center\subfigure[Daily
price]{\rotatebox{-0}{\includegraphics*[width=0.88\columnwidth]{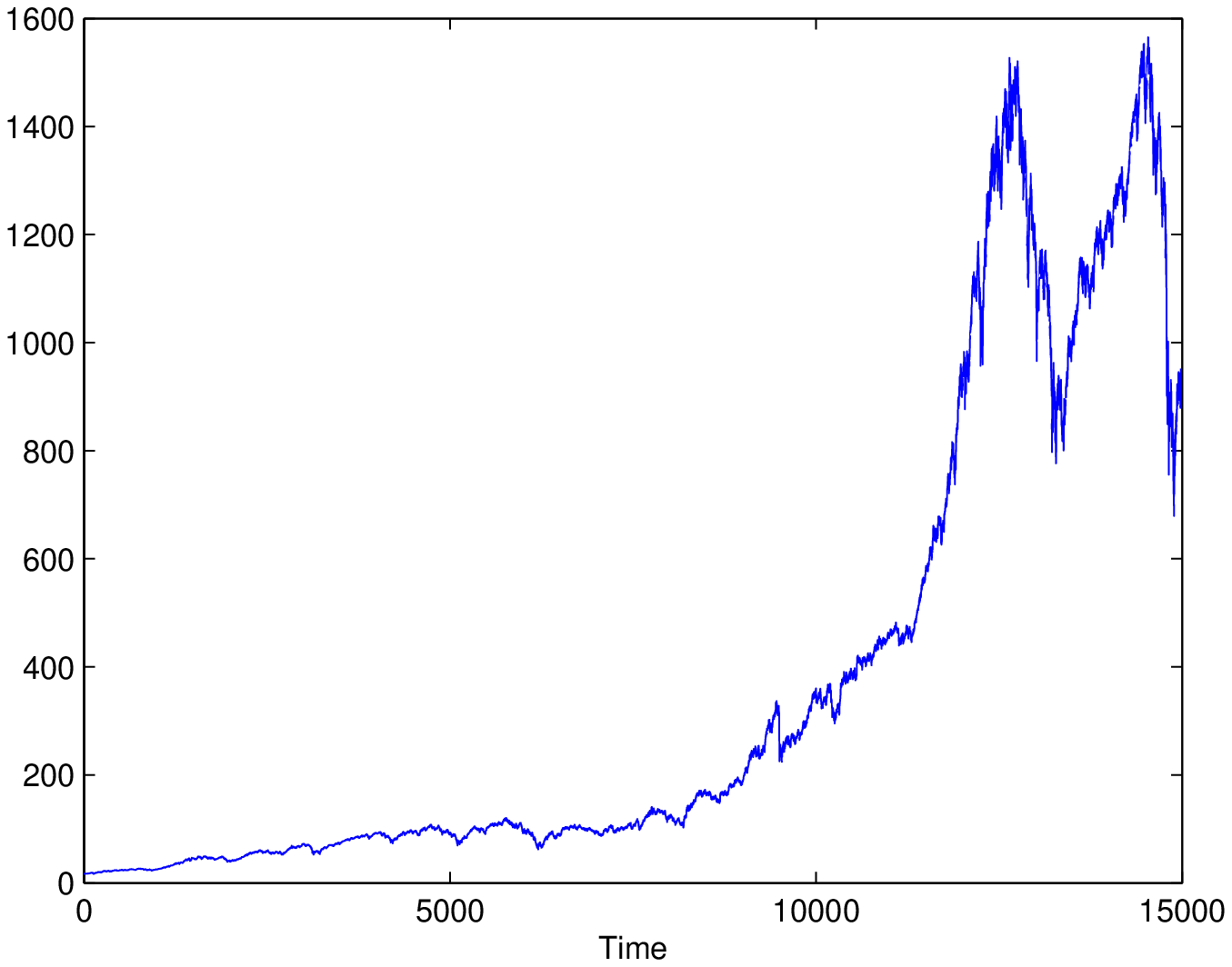}}}
\subfigure[Modified normalized logarithmic return
$r(t)$]{\rotatebox{-0}{\includegraphics*[width=0.88\columnwidth]{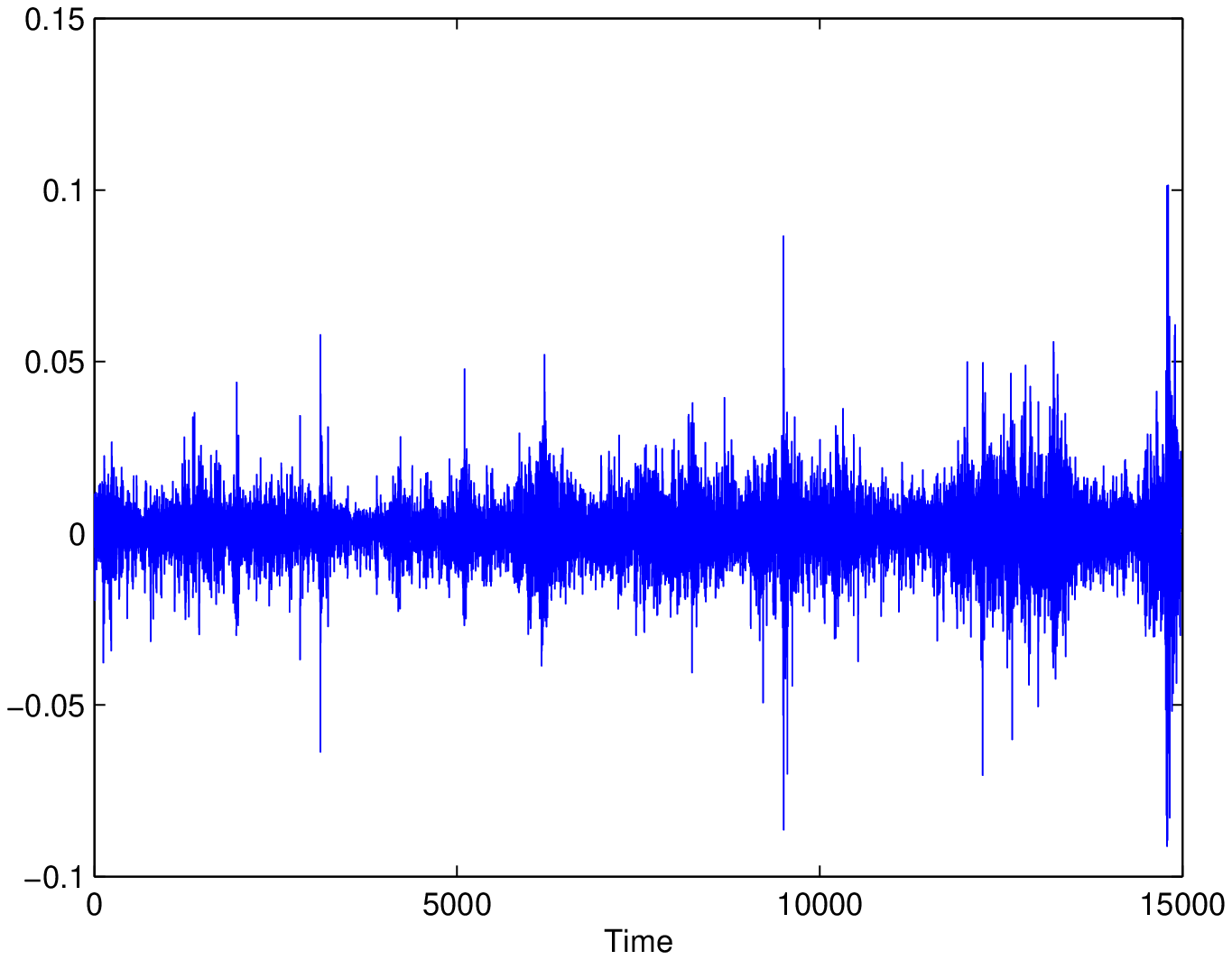}}}
\subfigure[Normalized mean logarithmic return $\bar
r(t)$]{\rotatebox{-0}{\includegraphics*[width=
0.88\columnwidth]{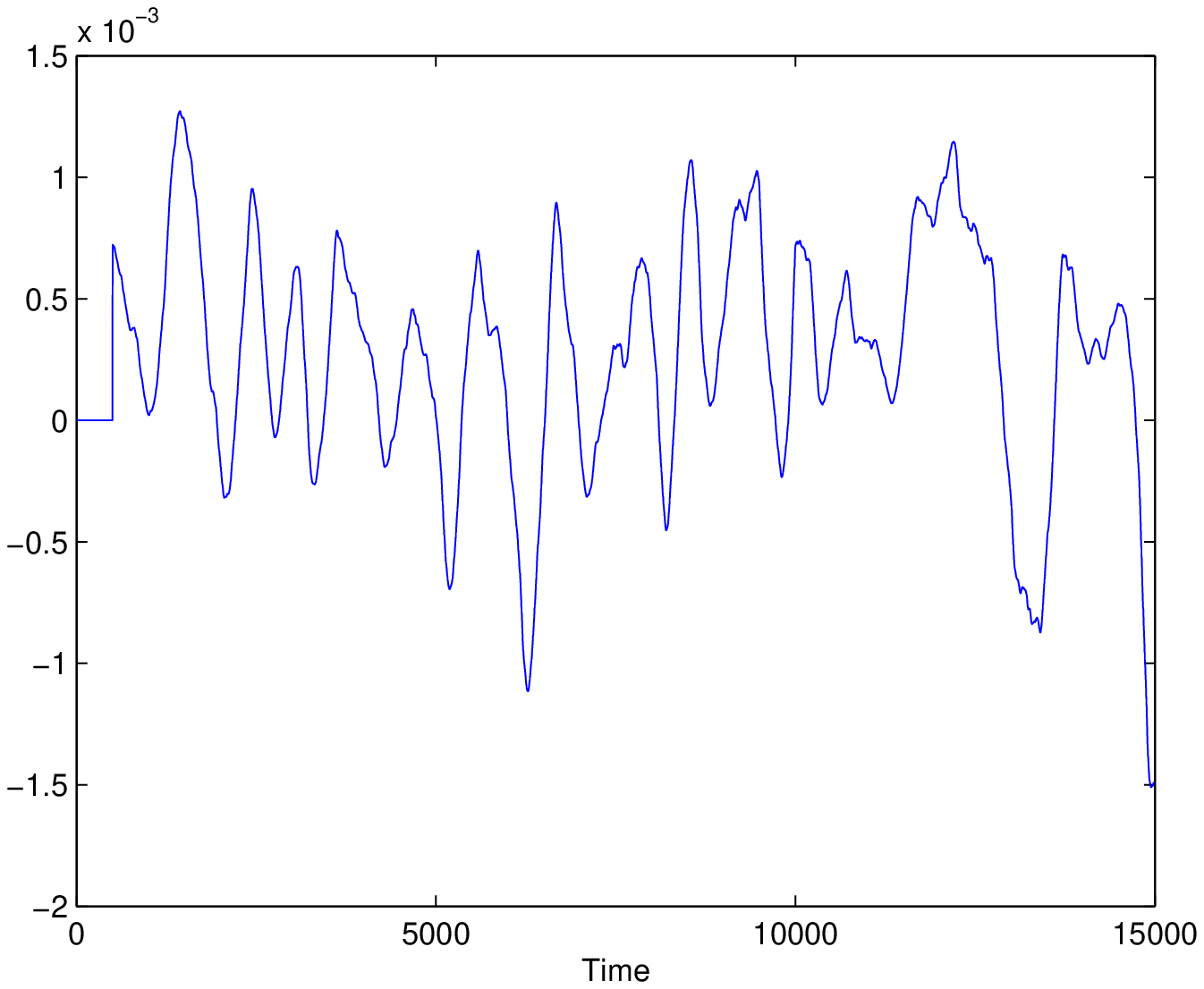}}}
\subfigure[$\text{\bf vol}({S\&P500})(t)$ (--) and 20 days
forecasting (- -)]{\rotatebox{-0}{\includegraphics*[width=
0.88\columnwidth]{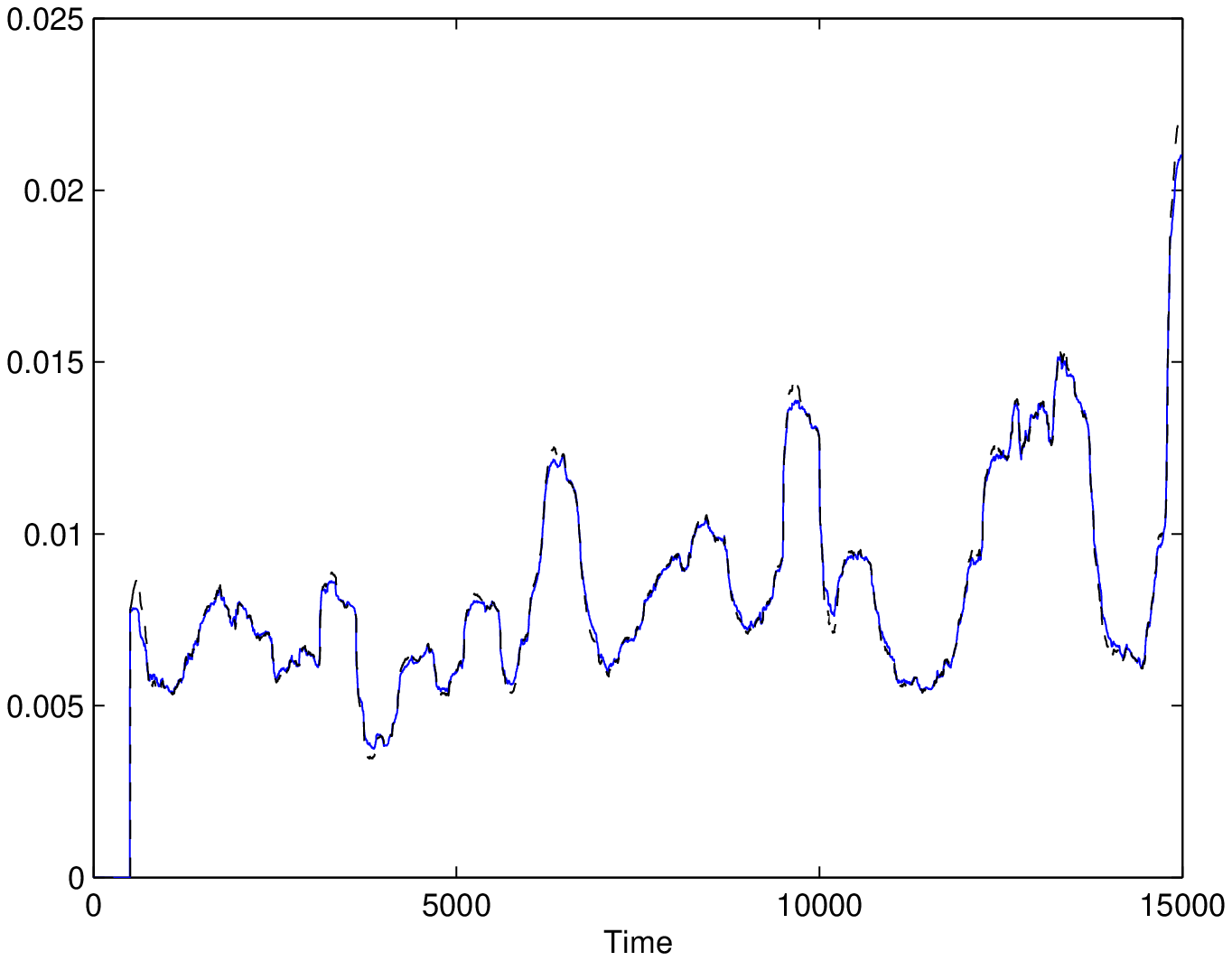}}}
 \caption{S\&P 500 \label{SP}}
\end{figure*}

\begin{figure*}
\center\subfigure[$\mathfrak{C}$: $\bar r_{IBM}(\bar r_{S\&P
500}$)]{\rotatebox{-0}{\includegraphics*[width=0.88\columnwidth]{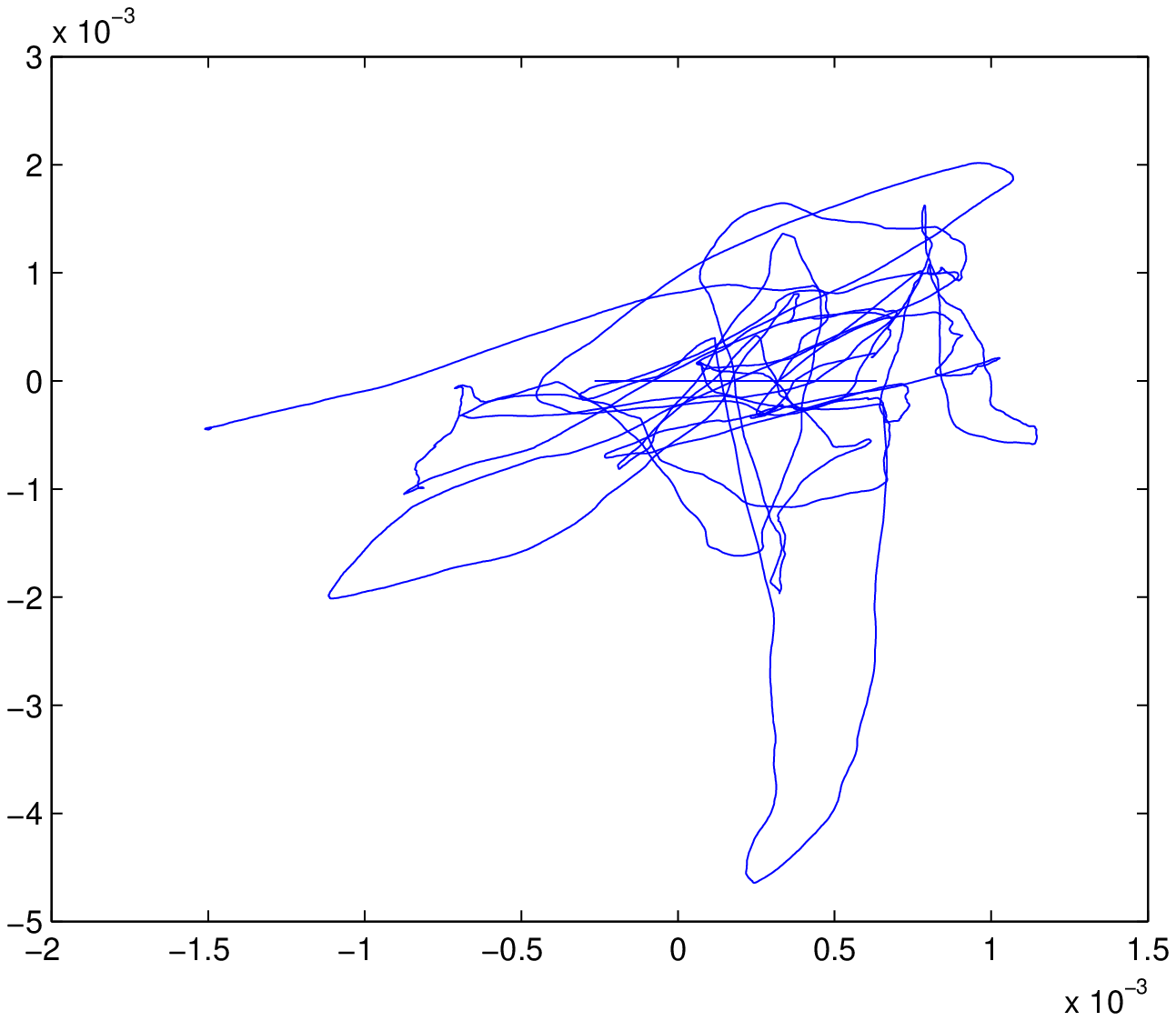}}}
\subfigure[IBM's
$\beta(t)$]{\rotatebox{-0}{\includegraphics*[width=0.88\columnwidth]{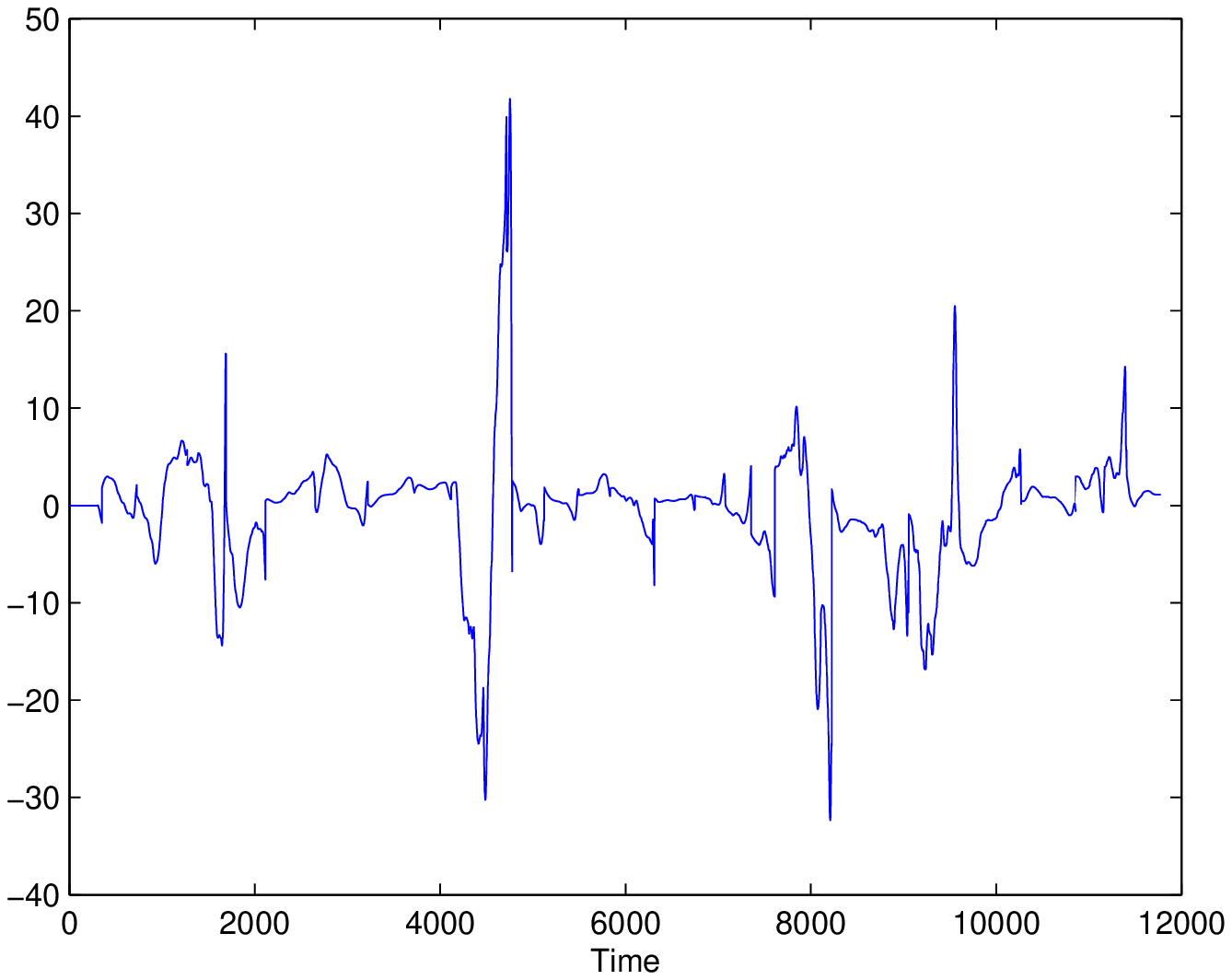}}}
\center\subfigure[$\mathfrak{C}$: $\bar r_{JPM}(\bar r_{S\&P
500})$]{\rotatebox{-0}{\includegraphics*[width=0.88\columnwidth]{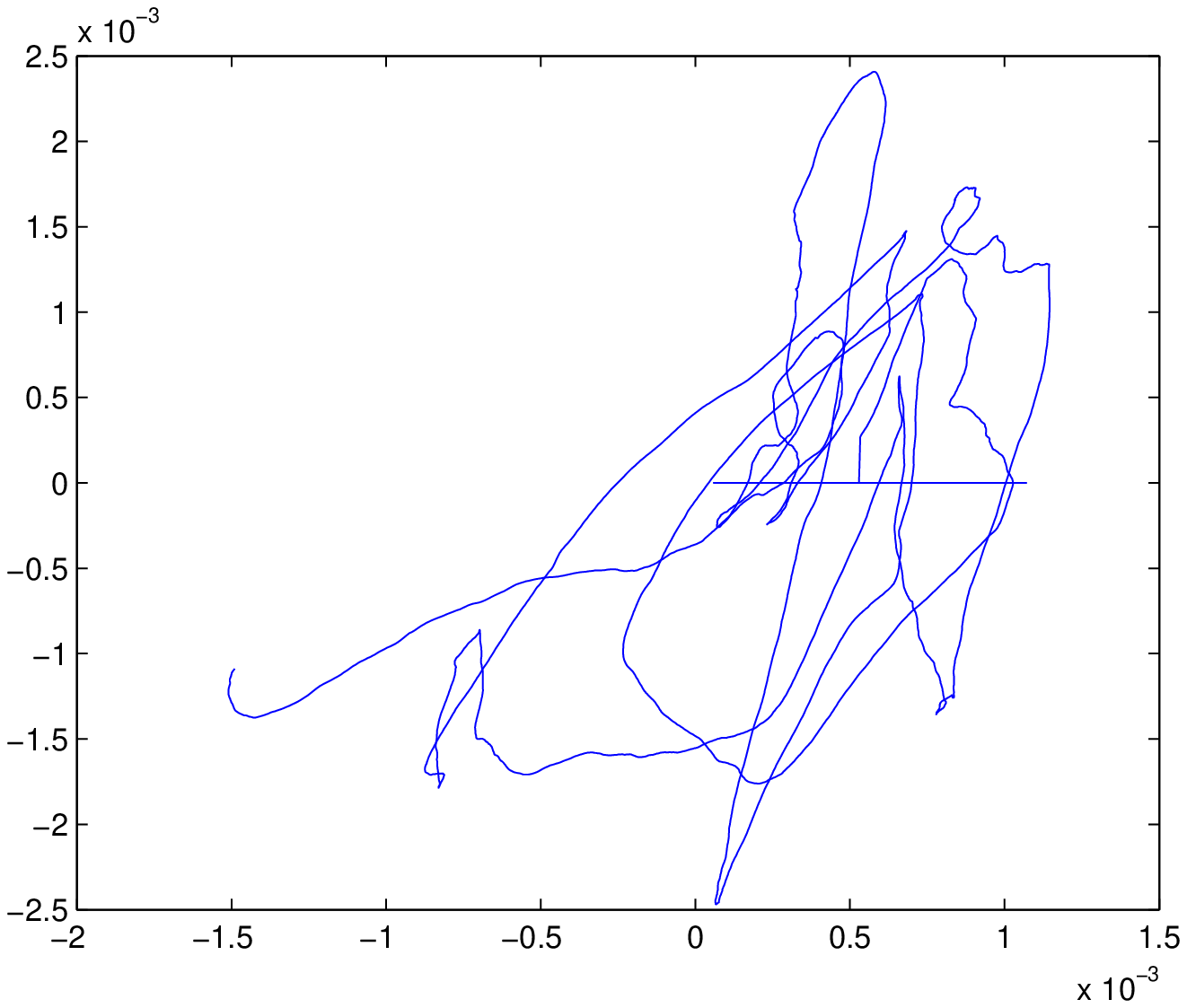}}}
\subfigure[JPM's
$\beta(t)$]{\rotatebox{-0}{\includegraphics*[width=0.88\columnwidth]{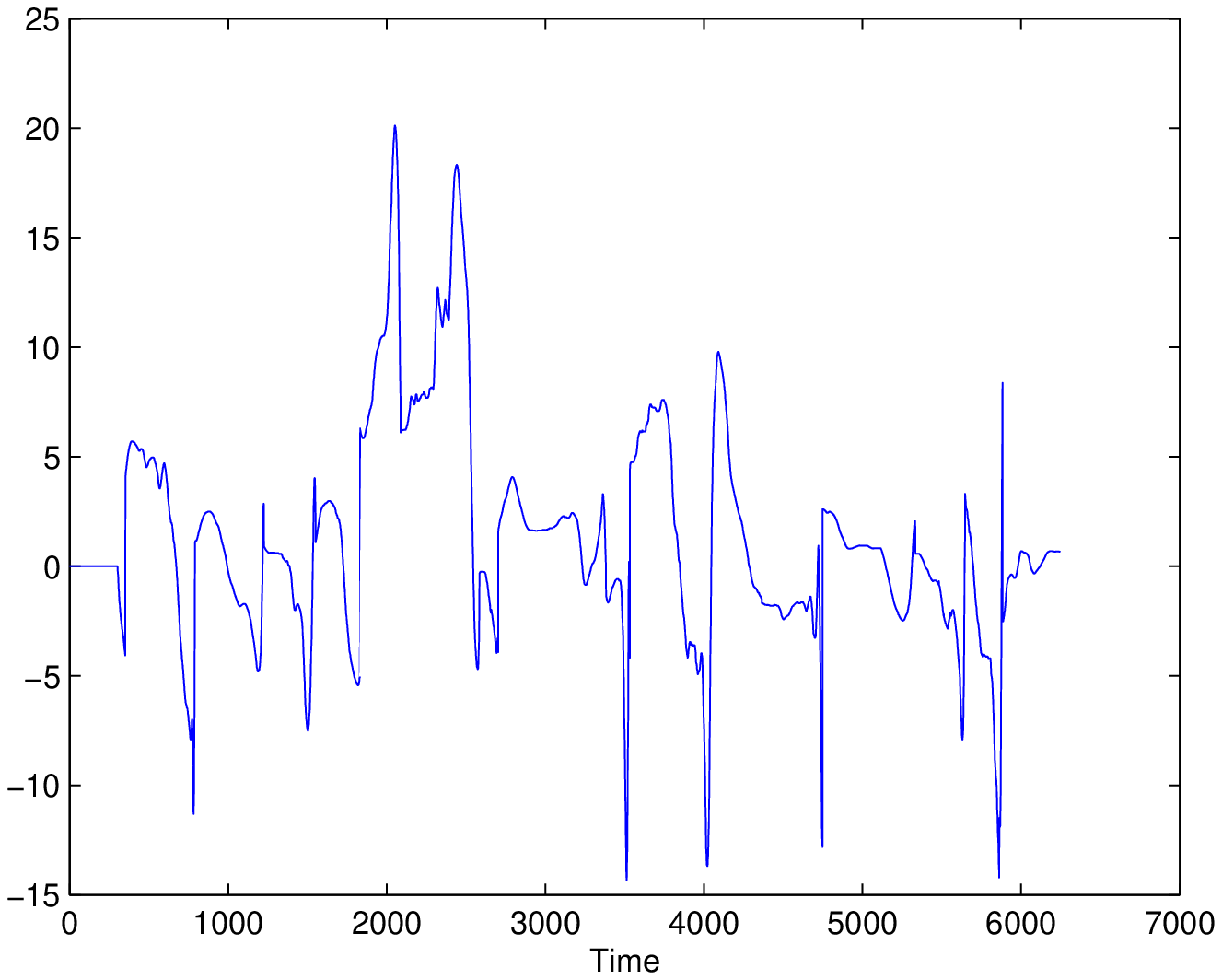}}}
%
%
 \caption{ $$\label{beta}}
\end{figure*}

The beta coefficients is computed with respect to the S\&P 500 (see
Figures \ref{SP}). The results displayed in Figures \ref{beta} are
obtained via the numerical techniques of \cite{beta}.

\begin{figure*}
\subfigure[$\text{SR}_{10}({S\& P 500})$]{\rotatebox{-0}{\includegraphics*[width=0.7\columnwidth]{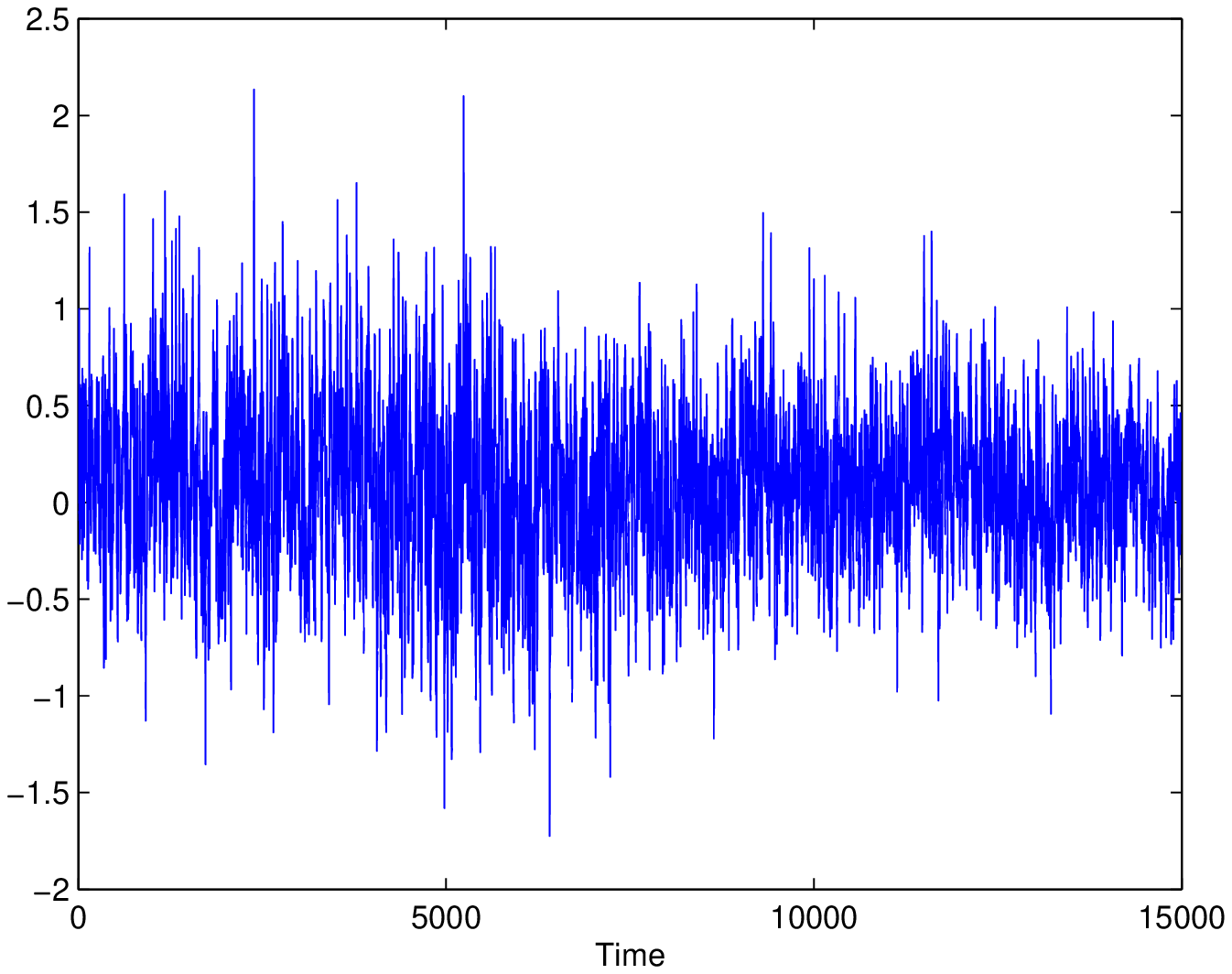}}}
\subfigure[$\text{SR}_{100}({S\& P 500})$ (--) and 10 days forecasting (- -)]{\rotatebox{-0}{\includegraphics*[width=0.7\columnwidth]{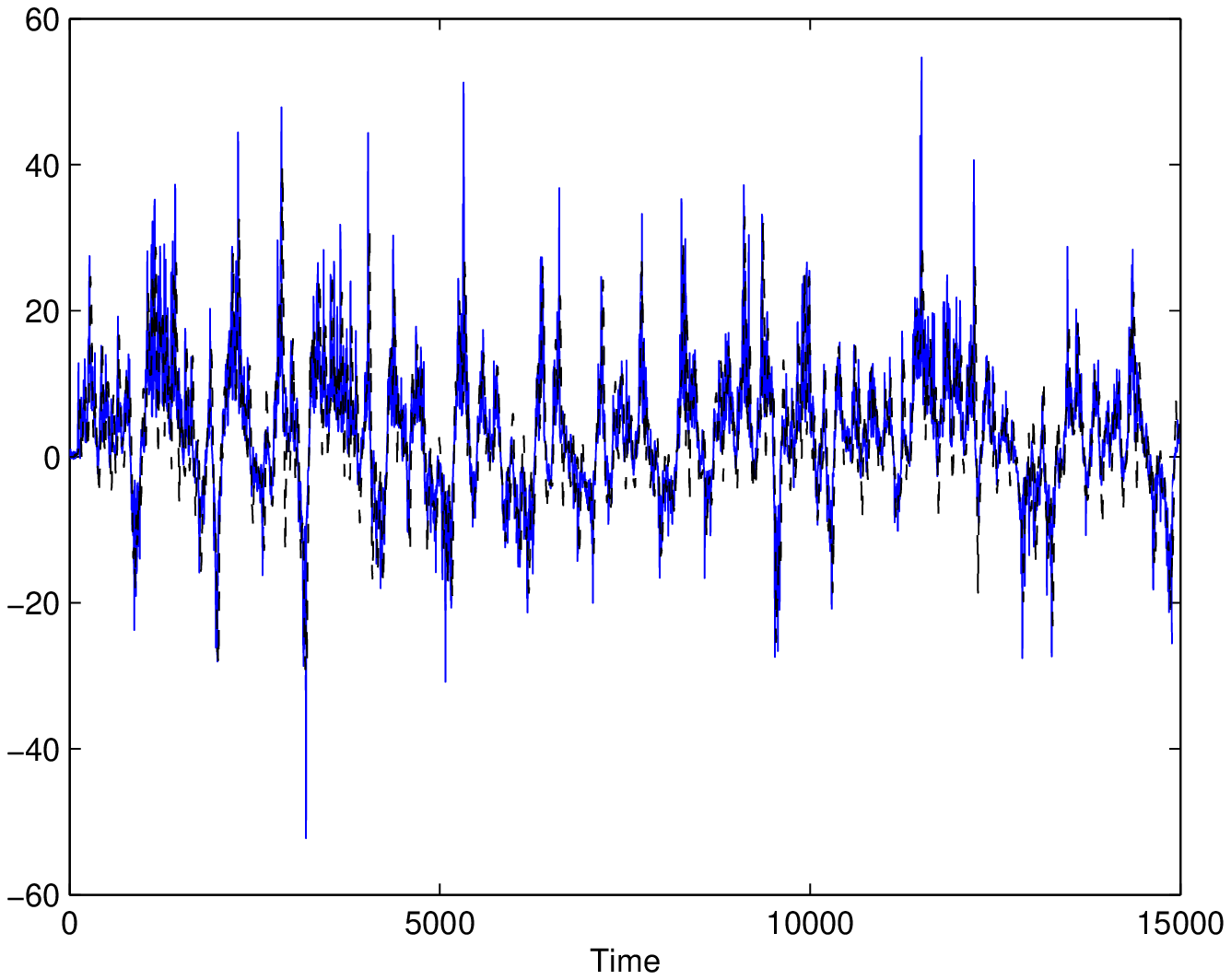}}}
\subfigure[Zoom of \ref{shr}-(b)]{\rotatebox{-0}{\includegraphics*[width=0.7\columnwidth]{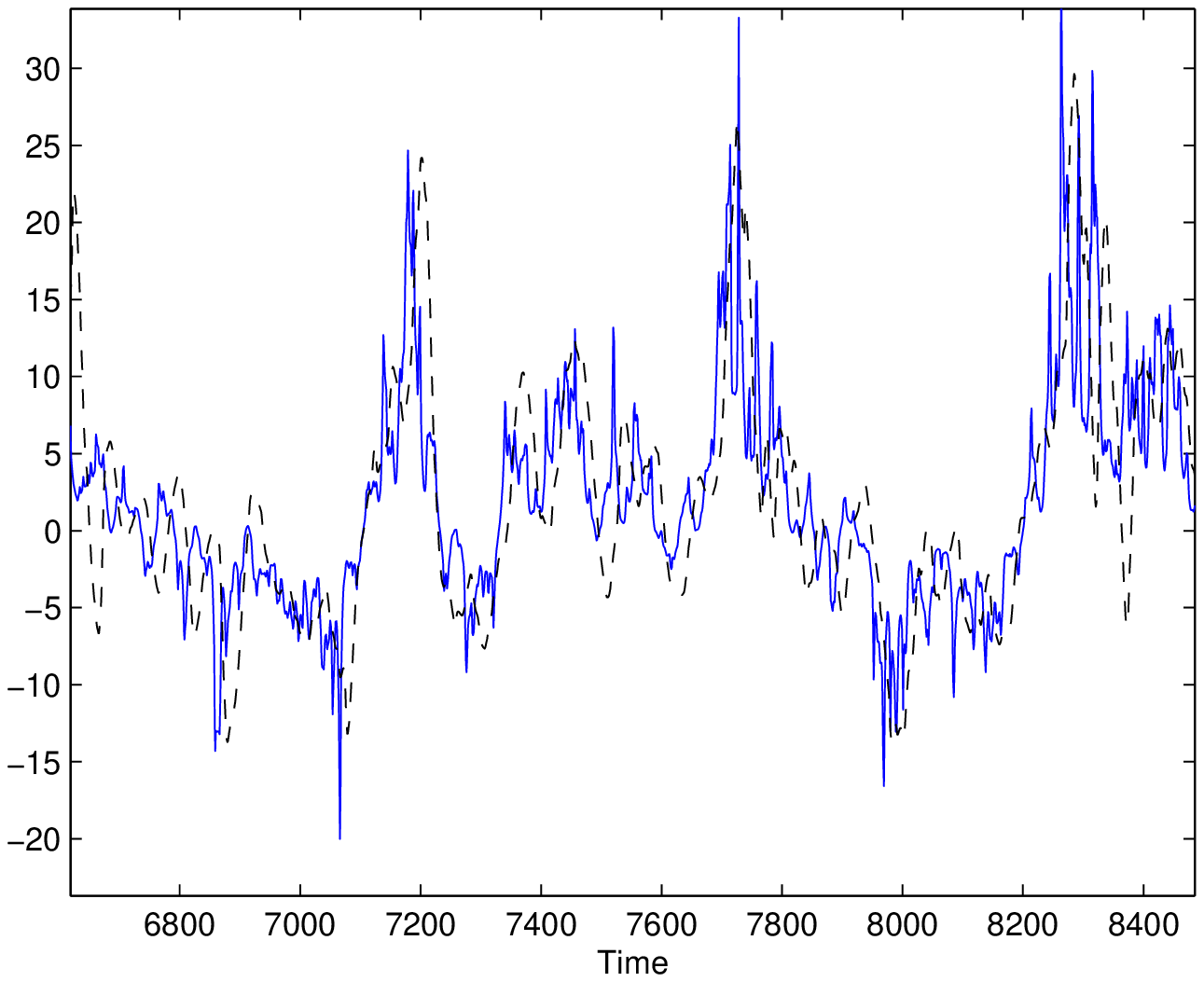}}}
\caption{ \label{shr}}
\end{figure*}

Figure \ref{shr} displays the Sharpe ratio of S\&P 500. With $\Delta
t=10$ a trend is difficult to guess in Figure \ref{shr}-(a). Figure
\ref{shr}-(b) on the other hand, where $\Delta t=100$, exhibits a
well-defined trend which yields a quite accurate forecasting of 10
days.



\section{Conclusion}\label{conclusion}
Although we have proposed a precise and elegant mathematical
definition of volatility, which \begin{itemize}
\item yields efficient and easily implementable computations,
\item will soon be exploited for a dynamic portfolio management \cite{agadir},
\end{itemize}
the harsh criticisms against its importance in financial engineering
should certainly not be dismissed (see, \textit{e.g.},
\cite{taleb}). Note for instance that we have not tried here to
forecast extreme events, \textit{i.e.}, abrupt changes (see
\cite{abrupt}) with this tool. This aim has been already quite
successfully achieved in \cite{fes,malo2,beta,delta}, not via
volatility but by taking advantage of indicators that are related to
prices and not to returns.


\end{document}